\documentclass[aps,pra,reprint,superscriptaddress,amsmath,amssymb,floatfix,longbibliography]{revtex4-2}
\usepackage{bm,mathtools}
\usepackage{graphicx,booktabs}
\usepackage{array}[=2016-10-06]
\usepackage{tabularx}
\usepackage{algpseudocode}

\makeatletter
\newcounter{algorithm}
\def\thealgorithm{\arabic{algorithm}}
\def\fps@algorithm{tbp}
\def\ftype@algorithm{4}
\def\ext@algorithm{loa}
\def\fnum@algorithm{Algorithm~\thealgorithm}
\newenvironment{algorithm}[1][tb]{\@float{algorithm}[#1]\small}{\end@float}
\newenvironment{algorithm*}[1][t]{\@dblfloat{algorithm}[#1]\small}{\end@dblfloat}
\makeatother
\usepackage{xcolor,tikz}
\usetikzlibrary{arrows.meta}
\usepackage{microtype}
\usepackage[hidelinks]{hyperref}
\algrenewcommand\algorithmicindent{1em}

\makeatletter
\newenvironment{widefigure}{\par\onecolumngrid\noindent\begin{minipage}{\textwidth}\def\@captype{figure}\centering}{\end{minipage}\par\vspace{12pt}\twocolumngrid}
\newenvironment{widetable}{\par\onecolumngrid\noindent\begin{minipage}{\textwidth}\def\@captype{table}\centering}{\end{minipage}\par\vspace{12pt}\twocolumngrid}
\makeatother

\usepackage{etoolbox}

\makeatletter
\newcommand{\separateauthornotes}[1]{%
  \begingroup
  \def\@join##1##2##3{%
    \@if@empty{##2}{}{##2\textsuperscript{,\,}}%
    \frontmatter@footnote{##3}%
  }%
  #1%
  \endgroup
}
\patchcmd{\doauthor}
  {\frontmatter@footnote{#2}}
  {\separateauthornotes{#2}}
  {}
  {\PackageError{authornotes}{REVTeX author-note patch failed}{}}
\makeatother

\begin{document}
\title{Warm-Start Iterative QITE for Distribution Network Reconfiguration via Branch-Exchange Encoding}
\author{Noah Crum}
\thanks{These authors contributed equally to this work.}
\email{crumn@epbqn.net}
\affiliation{EPB Quantum, Chattanooga, TN}

\author{Pete Pritchard}
\thanks{These authors contributed equally to this work.}
\email{pritchardp@epbqn.net}
\affiliation{EPB Quantum, Chattanooga, TN}

\author{Tanner Rase}
\thanks{These authors contributed equally to this work.}
\email{raset@epbqn.net}
\affiliation{EPB Quantum, Chattanooga, TN}
\author{Phillip C. Lotshaw}
\affiliation{Oak Ridge National Laboratory, Oak Ridge, TN}
\thanks{This manuscript has been authored by UT-Battelle, LLC, under Contract No. DE-AC0500OR22725 with the U.S. Department of Energy. The United States Government retains and the publisher, by accepting the article for publication, acknowledges that the United States Government retains a non-exclusive, paid-up, irrevocable, world-wide license to publish or reproduce the published form of this manuscript, or allow others to do so, for the United States Government purposes. The Department of Energy will provide public access to these results of federally sponsored research in accordance with the DOE Public Access Plan (\url{http://energy.gov/downloads/doe-public-access-plan}).}
\author{Shaked Regev}
\affiliation{Oak Ridge National Laboratory, Oak Ridge, TN}
\author{Claudio Girotto}
\affiliation{IonQ Inc., College Park, MD}
\author{Ananth Kaushik}
\affiliation{IonQ Inc., College Park, MD}
\author{Paul Smith}
\affiliation{EPB Quantum, Chattanooga, TN}

\begingroup
\renewcommand{\thefootnote}{}
\footnotetext{\textbf{Author contributions.}\enspace
N.C., P.P., T.R., P.L., and S.R. conceptualized the project.
N.C., P.P., and T.R. developed the theoretical framework and the circuit design.
N.C., P.P., T.R., C.G., and A.K. oversaw the experimental implementation on IonQ systems.
N.C., P.P., and T.R. performed numerical simulations relevant to the experiment, developed software toolkits, and performed data analysis.
N.C., P.P., T.R., P.L., and A.K. contributed to interpretation of results.
P.L., C.G., and P.S. supervised experiments and provided resources.
N.C., P.P., and T.R. produced all figures and drafted the manuscript.
All authors reviewed the manuscript.
}

\begin{abstract}
We present a hybrid quantum-classical algorithm for distribution network reconfiguration, a combinatorial optimization problem on power distribution networks, that combines a radiality-preserving branch-exchange encoding with iterative warm-start quantum imaginary-time evolution to minimize active line losses. The encoding uses a fixed-width binary representation of sequential branch-exchange actions, ensuring that every register outcome decodes to a radial configuration. The quantum subroutine is informed by a surrogate model fit to classically-solved alternating current power flow (ACPF) labels. After a set budget of ACPF solves and circuit samples is exhausted, the algorithm returns the lowest-loss configuration seen. We employ our algorithm on ten test cases, including seven commonly used benchmark cases and three cases we modified from these and similar standard systems. We include two methods to reduce problem size, which enable extensions of the algorithm to larger networks. We successfully reach minimal-loss configurations for six cases and find configurations with losses from $0.2\%-2.18\%$ above the reported reference minimal loss for the other four.
\end{abstract}

\maketitle

\section{Introduction}
Electrical grids are composed of generating stations, transmission systems, and distribution networks. Power produced by generating stations is carried through transmission systems to distribution networks for final delivery to consumers. Distribution networks usually possess more power transmission lines than are in use at any given time; faults or other events can disrupt the network or cause lines to fail. Normally-closed or sectionalizing switches located on some lines can be opened to electrically isolate faulted areas during repairs, and normally-open or tie switches on other lines can then be closed to prevent the isolation of unaffected areas of the network. Distribution networks are typically operated in radial configurations to simplify fault protection and coordination \cite{stevenson1982}, and must satisfy operational and engineering constraints for their safe and reliable function. Subject to these constraints, distribution networks may also be reconfigured in pursuit of arbitrary objectives, such as the minimization of power losses. The resulting combinatorial optimization problem on network configurations is known as distribution network reconfiguration (DNR).

DNR with a loss minimization objective seeks the radial network configuration that minimizes total active line losses, i.e., the real power dissipated as heat due to impedance as currents flow through the lines of the network. For a network with a single source, radial operation requires a connected tree. For a multi-source system, radial configurations are forests with one source per connected component of the network. Finding a solution to DNR is complicated by its status as a mixed-integer nonlinear optimization problem: it relies not only on continuous degrees of freedom such as voltages and currents, but also on binary variables that characterize the status of the switches. Moreover, determining the electrical data for the network given a specific configuration of switches requires the solution of nonlinear alternating current power flow equations (ACPF). Serious attention to DNR was first paid by Merlin and Back in 1975 \cite{merlin1975}, with notable later contributions by Shirmohammadi and Hong \cite{25637} and by Baran and Wu \cite{25627}, resulting in a well-developed field of classical computing methods for DNR today \cite{Lotfi2024, 193906, 19265, 1564201, 1208393, ZHANG2007685, CASTRO1985155, 667402, taylor_convex, CHANG1994227, Bharath_2025, ZHU200237, 6140618, 1645224, khodabaksh, hartmann2025}, including simulated annealing \cite{CHANG1994227}, genetic algorithms \cite{ZHU200237}, differential evolution \cite{1208393}, tabu search \cite{ZHANG2007685}, and convex relaxations \cite{taylor_convex, 6140618}. There is also a developing corpus of quantum computing approaches to address DNR \cite{GOLESTAN2023584, silva2023, 9919400, kaseb2024, en19051148, devalk2026} and power flow problems \cite{pnnl_2571592, 9423668, 10144277, gao2022solvingdcpowerflow, eskandarpour2021, hafshejani2024}. Quantum DNR formulations differ in how they represent topology, electrical losses, and constraints. Silva \emph{et al.} encode network constraints as QUBO penalties ~\cite{9919400, silva2023}, whereas de Valk \emph{et al.} retain higher-order terms to avoid quadratization~\cite{devalk2026}. Penalty-free approaches also exist: Hartmann \emph{et al.} construct quantum mixers that preserve the spanning-tree subspace~\cite{hartmann2025}. Bosmediano \emph{et al.} instead use a fixed mapping from tie closures to sectionalizing-switch openings and construct a QUBO from ACPF evaluations of all configurations of five tie-status bits~\cite{en19051148}. Quantum power flow methods address a separate task, including quantum linear solvers within decoupled power flow iterations~\cite{9423668, 10144277} and discrete formulations for quantum and digital annealers~\cite{kaseb2024}.

In this work, our central contributions are a hybrid quantum-classical algorithm to solve DNR and a fixed-width branch-exchange encoding that enforces radiality without penalties or topology-preserving mixers. Building on classical sequential and loop-based representations~\cite{braz2011sequential, barbosa2013codification, wen2016encoding, kim2023loop}, we represent switching decisions as ordered actions applied to a radial default. Each action either leaves the current configuration unchanged or closes a tie switch and opens another switch on the resulting cycle. Under this approach, every sampled action tuple therefore decodes to a radial configuration, providing an interface between quantum candidate generation and classical ACPF evaluation in a surrogate-assisted search for lower active line losses.

The resulting workflow separates candidate selection from electrical evaluation, using quantum sampling and a learned-loss model to guide the search while retaining classical ACPF to assess each selected configuration. A converged ACPF solve with finite loss supplies a usable label, consisting of the configuration and its evaluated loss. Following established surrogate-assisted optimization approaches~\cite{kitai2020fmqa, baptista2018bocs}, we fit a second-order model of the branch-exchange actions to these labels. This surrogate predicts losses for candidate selection, whereas the ACPF evaluations determine the best observed configuration, termed the incumbent. Since different register states can describe the same configuration, both fitting and rank account explicitly for these repeated representations. 

To propose additional candidates, we map the fitted surrogate to a diagonal Hamiltonian and prepare a warm-start product state centered on a register representation of the incumbent configuration. The state initializes a parameterized QITE circuit based on an adaptation of the expectation-matching construction of Lotshaw \emph{et al.}~\cite{lotshaw2026}, with circuit parameters computed classically from analytic expectations of the initial product state under the fitted surrogate Hamiltonian.

After measurement, we decode the register states and merge repeated configurations before ranking distinct unevaluated candidates by their surrogate predictions averaged over their register representations. ACPF evaluations of the selected candidates then supply new labels for refitting the surrogate and updating the incumbent, completing the acquisition loop shown in Fig.~\ref{fig:workflow}. The search repeats this process within prescribed budgets for usable ACPF labels and circuit samples, returning the lowest-loss configuration evaluated. For Test Cases 1--5, simulation and hardware results achieve optimal loss configurations verified by ACPF over all usable radial configurations with this method.

Due to the limited qubit counts of quantum hardware, we introduce two approximate methods that limit register size to extend the workflow to larger distribution networks: a DISTOP-inspired switch-fixing technique \cite{25637} and a rolling-window method over subsets of tie switches. The first reduces the number of switchable branches; the second outlines a procedure to divide the network into smaller pieces that can be addressed in sequence. For the larger Test Cases 6--10 where full ACPF enumeration of feasible configurations is intractable, we employ the rolling-window method and compare our optimization results to those reported in the literature, approaching optimal results. For all instances, we track the best observed loss as distinct configurations are evaluated. 

The manuscript proceeds as follows: Section \ref{sec:methods} describes the methods including details of the branch-exchange encoding, construction of our second-order surrogate model, the expectation-matching evolution procedure with QITE, and extensions to larger network sizes, Section \ref{sec:results} presents the results of our procedure over ten network cases, and Section \ref{sec:conclusion} discusses conclusions and future work. The supplementary appendices provide fuller derivations and implementation details.

\begin{figure*}[!htbp]
    \centering
    \resizebox{\textwidth}{!}{%
\begingroup%
\providecommand{\QITEWorkflowTextFont}{\fontfamily{ptm}\selectfont}%
\definecolor{wfInk}{HTML}{292929}%
\definecolor{wfMuted}{HTML}{626262}%
\definecolor{wfBorder}{HTML}{A0A0A0}%
\definecolor{wfPurple}{HTML}{6F3C8A}%
\definecolor{wfPurplePale}{HTML}{F3EEF8}%
\definecolor{wfGrey}{HTML}{5F5F5F}%
\definecolor{wfGreyPale}{HTML}{F0F0F0}%
\definecolor{wfGrayPale}{HTML}{F7F7F7}%

\begin{tikzpicture}[
  x=1mm, y=-1mm,
  every node/.style={
    text=wfInk, anchor=west, inner sep=0pt, outer sep=0pt,
    font=\normalfont\QITEWorkflowTextFont\fontsize{9.5}{11.4}\selectfont
  },
  wfbox/.style={draw=wfBorder, fill=white, line width=0.8pt,
    rounded corners=1.1mm},
  wfhead/.style={font=\normalfont\QITEWorkflowTextFont\bfseries\fontsize{10}{12}\selectfont},
  wfsetuphead/.style={font=\normalfont\QITEWorkflowTextFont\bfseries\fontsize{10.6}{12.7}\selectfont},
  wfsetuptext/.style={font=\normalfont\QITEWorkflowTextFont\fontsize{10}{12}\selectfont},
  wfflow/.style={-{Stealth[length=1.7mm,width=1.25mm]},
    draw=wfMuted, line width=0.9pt},
  wfwire/.style={draw=wfPurple, line width=0.7pt},
  wfgate/.style={draw=wfPurple, fill=white, line width=0.75pt},
  wfgatetext/.style={anchor=center, text=wfPurple,
    font=\normalfont\fontsize{10.95}{13.14}\selectfont}
]
\path[use as bounding box] (0,0) rectangle (180,117);

\draw[draw=wfMuted, line width=0.8pt, line cap=round,
  dash pattern=on 0pt off 2.4pt, rounded corners=1.5mm]
  (2,37) rectangle (179,92);

\draw[wfflow] (86,16) -- (94,16);
\draw[wfflow] (134.5,28) -- (134.5,34) -- (24,34) -- (24,46);
\draw[wfflow,draw=wfPurple] (44,62) -- (49,62);
\draw[wfflow,draw=wfPurple] (88,62) -- (93,62);
\draw[wfflow] (132,62) -- (137,62);

\draw[wfbox,fill=wfGrayPale] (5,4) rectangle (86,28);
\node[wfsetuphead] at (7.5,9.5) {1\enspace Encode the network};
\node[wfsetuptext] at (7.5,17) {Specify the network and radial default};
\node[wfsetuptext] at (7.5,22.6) {Construct the branch-exchange register};

\draw[wfbox,fill=wfGrayPale] (94,4) rectangle (175,28);
\node[wfsetuphead] at (96.5,9.5) {2\enspace Initialize loss labels};
\node[wfsetuptext] at (96.5,17) {Evaluate initial configurations with ACPF};
\node[wfsetuptext] at (96.5,22.6) {Store usable losses and best configuration};

\draw[wfbox] (5,46) rectangle (44,78);
\node[wfhead] at (7.5,51.5) {3\enspace Fit the surrogate};
\node at (7.5,59) {Fit the loss model};
\node at (7.5,64.6) {Select warm start};
\node at (7.5,70.2) {Set QITE parameters};

\draw[wfbox,draw=wfPurple,fill=wfPurplePale,line width=1.05pt]
  (49,46) rectangle (88,78);
\node[wfhead,text=wfPurple] at (51.5,51.5) {4\enspace Sample the register};
\foreach \yy in {59,64,69} {
  \draw[wfwire] (51.5,\yy) -- (85.8,\yy);
}
\draw[wfgate] (53.5,55.8) rectangle (64.5,72.2);
\draw[wfgate] (67.5,55.8) rectangle (77.5,72.2);
\node[wfgatetext] at (59,64) {$|\psi_{0,r}\rangle$};
\node[wfgatetext] at (72.5,64) {$U(\bm\mu)$};
\foreach \yy in {59,64,69} {
  \draw[wfgate] (80.5,{\yy-1.7}) rectangle (84.9,{\yy+1.7});
  \draw[wfwire] (81.45,{\yy+0.45})
    .. controls (81.5,{\yy-1.2}) and (83.9,{\yy-1.2}) .. (83.95,{\yy+0.45});
  \draw[wfwire] (82.7,{\yy+0.45}) -- (83.6,{\yy-0.95});
}
\node[anchor=center,text=wfPurple] at (68.5,75) {Prepare, evolve, measure};

\draw[wfbox] (93,46) rectangle (132,78);
\node[wfhead] at (95.5,51.5) {5\enspace Select configurations};
\node[anchor=center,font=\normalfont\fontsize{12}{14.4}\selectfont]
  at (112.5,59.6)
  {$\bm{x}\;\longrightarrow\;\bm{a}
    \;\overset{\mathsf{Dec}}{\longrightarrow}\;\bm{z}$};
\node at (95.5,66.5) {Merge duplicates};
\node at (95.5,72.1) {Rank untried candidates};

\draw[wfbox,draw=wfGrey,fill=wfGreyPale] (137,46) rectangle (176,78);
\node[wfhead,text=wfGrey] at (139.5,51.5) {6\enspace Evaluate losses};
\node at (139.5,59) {Full-network ACPF};
\node at (139.5,64.6) {Add usable loss labels};
\node at (139.5,70.2) {Update best configuration};

\draw[draw=wfGrey,line width=0.9pt] (156,78) -- (156,88);
\draw[wfflow,draw=wfGrey] (156,88) -- (24,88) -- (24,78);
\fill[wfGrey] (156,88) circle[radius=0.45mm];
\draw[wfflow,draw=wfGrey] (156,88) -- (156,101);
\draw[wfbox,fill=wfGrayPale] (5,101) rectangle (175,113);
\node[anchor=center,font=\normalfont\QITEWorkflowTextFont\bfseries\fontsize{10.5}{12.6}\selectfont]
  at (90,107) {Return the configuration with the lowest observed ACPF loss};

\node[anchor=center,text=wfMuted,fill=white,inner xsep=3pt]
  at (80,34) {Begin acquisition};
\node[anchor=center,text=wfPurple] at (68.5,41.5) {Quantum circuit};
\node[anchor=center,text=wfGrey,fill=white,inner xsep=3pt]
  at (88,88) {Repeat steps 3--6 with updated loss labels};
\node[anchor=east,text=wfMuted,fill=white,inner xsep=3pt]
  at (152,96) {Stop at the budget or an exit condition};
\end{tikzpicture}%
\endgroup%
%
    }
    \caption{Surrogate-assisted quantum imaginary-time evolution (QITE) for distribution network reconfiguration using branch-exchange encoding. The workflow begins with a radial default configuration, from which the branch-exchange register is constructed, while initial ACPF evaluations establish the loss-label set and incumbent configuration. In each subsequent round, the surrogate is fitted to the available labels, and the warm-start and QITE parameters are determined classically. The quantum circuit then prepares a fresh warm-start state, applies one QITE update, and samples the register. Measured bits $\bm{x}$ are unpacked into actions $\bm a$ and decoded into configurations $\bm z$; aliases are merged, and configurations are ranked by surrogate prediction and the lowest-loss predictions according to the surrogate undergo ACPF evaluation. Converged evaluations with finite losses update the label set and incumbent, providing the data for the next round. This cycle continues until a budget or stopping condition is reached, after which the configuration with the lowest observed ACPF loss is returned. }
    \label{fig:workflow}
\end{figure*}
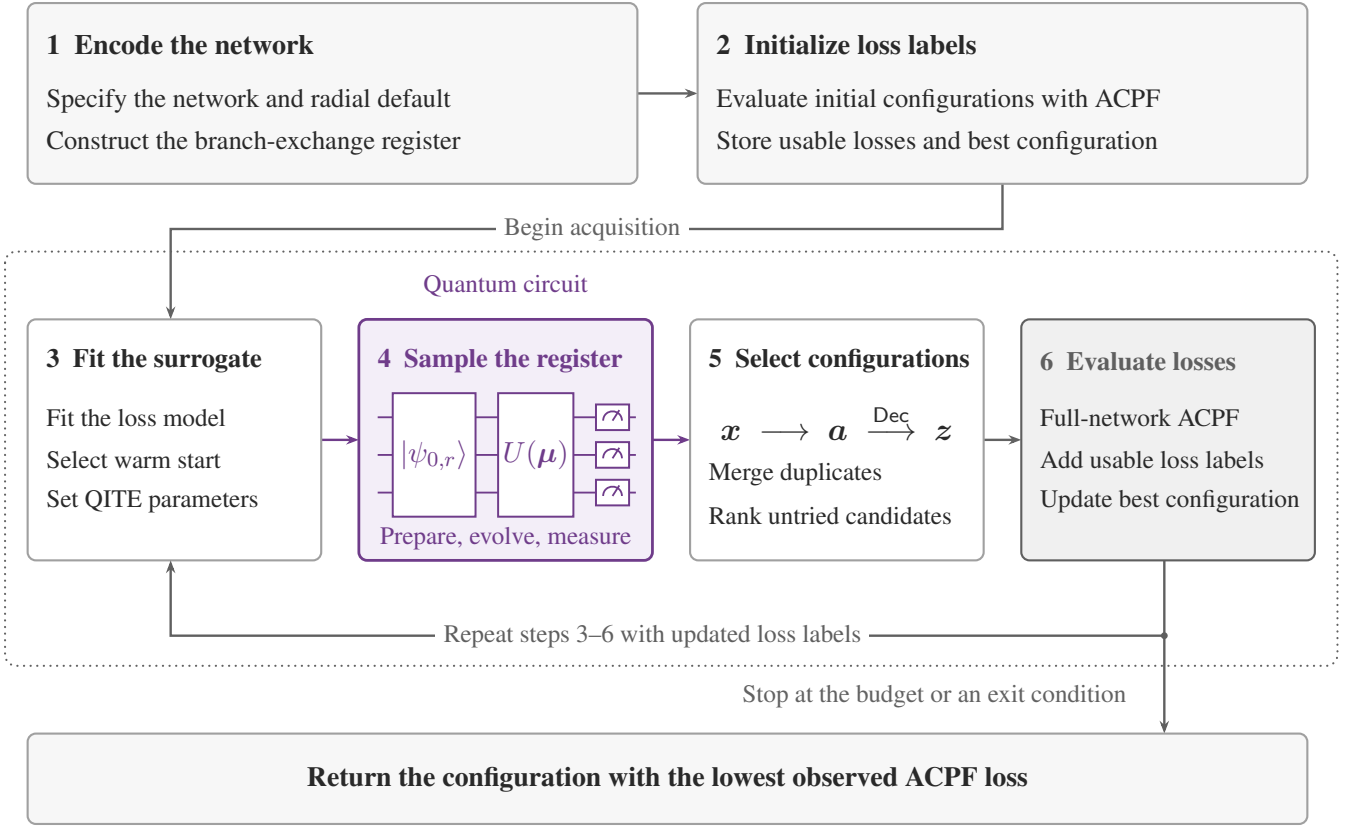

\section{Methods}\label{sec:methods}
We begin by defining the set of radial configurations. Let the graph $\mathcal G = (\mathcal V,\mathcal E_f \cup \mathcal E_s)$ describe a network, with buses $\mathcal V$, branches held closed $\mathcal E_f$, and controlled switches $\mathcal E_s=\{e_1,\ldots, e_m\}$. Branches held open do not enter the graph and are omitted. The switch vector $\bm z \in \{0,1\}^m$ of a configuration has $z_j=1$ when $e_j$ is closed and $z_j=0$ when $e_j$ is open. The graph $\mathcal G(\bm z)$ contains all buses, all fixed branches, and the closed controlled branches. With a set $S \subseteq \mathcal V$ of source buses, define
\begin{equation}
    \mathcal F=\left\{\bm z:\begin{array}{l}
    \mathcal G(\bm z)\text{ is a forest},\\
    |C\cap\mathcal S|=1\quad\forall C\in\operatorname{cc}(\mathcal G(\bm z))
    \end{array}\right\},
    \label{eq:radial_set}
\end{equation}
where $\operatorname{cc}(\cdot)$ returns connected components and $C$ denotes a component's bus set. Eq. ~\eqref{eq:radial_set} defines the radial set and excludes source-free islands and connections between distinct sources. We focus on meeting the radiality constraint; other operational constraints of ACPF are not addressed by this set of feasible configurations.

We solve ACPF using the DSS-Extensions implementation of OpenDSS~\cite{opendss2011}, accessed through the OpenDSSDirect Python interface~\cite{opendssdirect2024}. For each configuration $\bm z$, the evaluator applies the specified switch states to the network and performs a snapshot ACPF solve. It returns total active losses $L(\bm z)$ in kW, and a label is usable only if the solve converges and the loss is finite. Ultimately, the search seeks the lowest usable loss. Table~\ref{tab:notation} summarizes the notation used throughout this work.

\begin{table}[!htbp]
\caption{Principal notation}
\label{tab:notation}\centering\small
\begin{tabularx}{\columnwidth}{@{}lX@{}}
\toprule
Symbol & Meaning\\\midrule
$m,p,q$ & Numbers of controlled switches, tie switches, and qubits\\
$\bm z$ & Switch vector; one bit per controlled switch\\
$\bm a, \bm x$ & Action tuple and binary representation of action tuple\\
$a_s,w_s,K_s$ & Stage action, bit width, and the size of the action space $K_s=2^{w_s}$\\
$\mathsf{Dec},\mathcal O(\bm z)$ & Decoder and complete alias set of configuration $\bm z$\\
$\mathcal D_r,n_r$ & Usable labels before round $r$ and their count\\
$N,N_0,B,S_r$ & Label cap, requested initial labels, batch size, and shots in round $r$\\
\bottomrule
\end{tabularx}
\end{table}

\subsection{Sequential Branch-Exchange Encoding}\label{sec:encoding}
We define a default configuration with switch vector $\bm z^{(0)}\in\mathcal F$ for each network. Additionally, we consider the fixed order $(t_0,\ldots,t_{p-1})$ of its $p$ open ties which defines the sequentially processed stages $s$ of the encoding. Each tie specifies a single stage for the decoder. For distribution systems with multiple sources, connect each source to a fictitious super-root via a fixed edge. This modification is made for the topological analysis to follow; the added edges have no effect on ACPF verification and cannot be opened by branch-exchange actions.

Fundamentally, the branch-exchange encoding represents switching decisions by an action tuple $\bm a=(a_0,\ldots,a_{p-1})$. At stage $s$, action $a_s$ applies to tie $t_s$ of the configuration produced by the preceding stages, and is an integer greater than or equal to zero. The action value represents an index in the list of edges of the cycle that results when the tie $t_s$ is closed and indicates which edge to open to break the cycle and restore radiality. Action zero leaves this configuration unchanged, while a positive action closes $t_s$ and selects a controlled branch to open.

To define the branch-selection order, first find the unique path between the endpoints of $t_s$ in the current tree. We choose the traversal direction by comparing the order of the endpoint buses, e.g. \texttt{bus2} precedes \texttt{bus10}. This convention determines which endpoint is visited first. The path is then traversed from that endpoint to the other, retaining the controlled branches in the order encountered and omitting the fixed branches. The resulting list is: 
\begin{equation}
P_s(\bm z^{(s)})=(e_{s,1},\ldots,e_{s,\ell_s}).
\label{eq:stage_path}
\end{equation}
Any fixed branches in this traversal participate in connectivity but they do not appear in the path. The path and its length $\ell_s$ are stage dependent as subsequent branch-exchange actions impact the cycle-structure of the graph.

Let $\mathcal B_s$ be the set of closed branches before stage $s$, including any fixed and super-root edges. For a branch-exchange action on stage $s$, $a_s \in \{0,\ldots,K_s-1\}$ where $K_s$ denotes the number of actions supported by the bit register allocated to that stage, set
\begin{align}
j_s&=1+((a_s-1)\bmod\ell_s),\qquad a_s>0, \label{eq:selection}\\ 
\mathcal B_{s+1}&=\begin{cases}
\mathcal B_s,&a_s=0,\\
(\mathcal B_s\cup\{t_s\})\setminus\{e_{s,j_s}\},&a_s>0.
\end{cases}
\label{eq:decoder_step}
\end{align}
Zero here serves as a skip or no-operation signifier. If $a_s=0$, leave the branch set $\mathcal B_s$ unchanged. If $a_s>0$, close that stage's tie $t_s$ and open the controlled switch on the induced cycle indicated by the exchange action according to Eq.~\eqref{eq:selection}. Branch-exchange actions that exceed the length of the currently considered path wrap around in the modular arithmetic of Eq. ~\eqref{eq:selection}.
\begin{algorithm}[t]
\caption{Decode one register state}\label{alg:decoder}
\begin{algorithmic}[1]
\Require Radial default, ordered ties, widths $w_s$, register bits $\bm x$
\State Unpack $\bm x$ into stage actions $\bm a$
\State $B\gets$ closed default branches and super-root edges
\For{$s=0,\ldots,p-1$}
\If{$a_s>0$}
\State Find the current ordered opening list $P_s$
\State $j\gets1+((a_s-1)\bmod|P_s|)$
\State $B\gets(B\cup\{t_s\})\setminus\{P_s[j]\}$
\EndIf
\EndFor
\State Remove super-root edges; return switch vector $\bm z$
\end{algorithmic}
\end{algorithm}

\emph{Branch-Exchange quantities.} Let $\mathcal R_s$ denote the distinct switch vectors reachable before stage $s$, with $\mathcal R_0=\{\bm z^{(0)}\}$. For each $\bm z\in\mathcal R_s$, let $\ell_s(\bm z)$ be the number of eligible switches on the current path between the endpoints of tie $t_s$. A forward pass constructs these reachable states and determines
\begin{equation}
M_s=\max_{\bm z\in\mathcal R_s}\ell_s(\bm z).
\label{eq:max_opening_length}
\end{equation}
To construct the register, we must represent every possible action over the processed path. Therefore, we define the bit width of a stage accordingly. Furthermore, we define offset values to indicate the separation of stage allocations with respect to register bits:
\begin{equation}
w_s=\lceil\log_2(1+M_s)\rceil,\quad
o_s=\sum_{h<s}w_h,\quad q=\sum_{s=0}^{p-1}w_s.
\label{eq:widths}
\end{equation}
The additional action in Eq.~\eqref{eq:widths} represents the pass operation $a_s=0$. These are the smallest fixed widths that retain every local choice in this stage layout. For our register, qubit zero is the least-significant bit, and the basis ket is $|x_{q-1}\cdots x_0\rangle$. Algorithm~\ref{alg:decoder} implements Eqs.~\eqref{eq:stage_path}--\eqref{eq:decoder_step}. See Fig.~\ref{fig:case-1-example} for an illustrative example of the branch-exchange encoding for Case 1 of our study.

\begin{figure*}[!htbp]
\centering
\includegraphics[width=\textwidth]
    {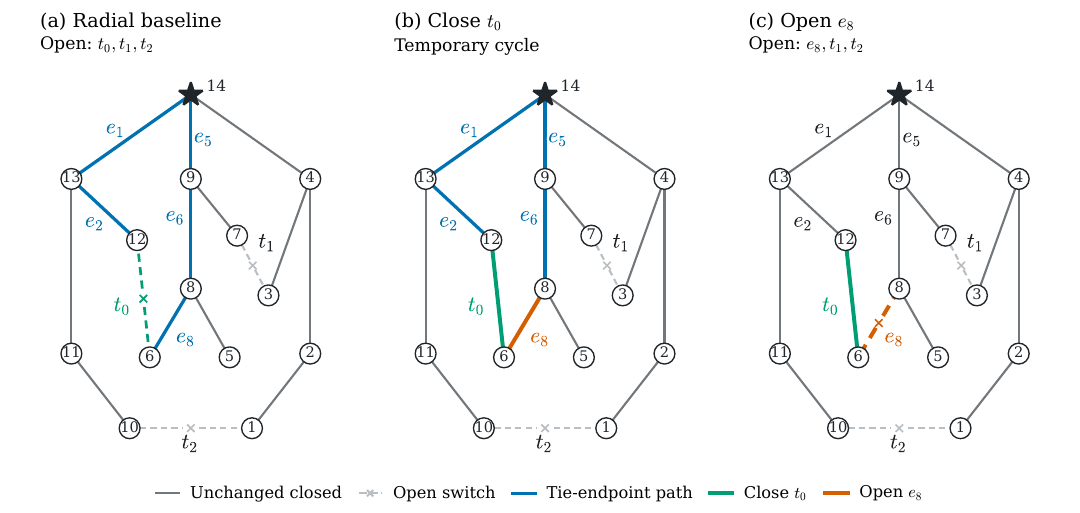}
\caption[Branch-exchange example for Case-1]{Branch-exchange encoding for Test Case 1. The ordered ties $(t_0,t_1,t_2)$ require stage widths $(3,4,4)$, giving an 11-qubit register with $2^{11}=2{,}048$ branch-exchange codes that decode to 190 distinct radial configurations. (a) The default configuration is radial; the star marks the source bus. (b) Closing $t_0$ forms one temporary cycle.
The ordered path between its endpoints is $P_0=(e_8, e_6, e_5, e_1, e_2)$. (c) Opening $e_8$ restores radiality. The action tuples $(1,0,0)$ and $(6,0,0)$ both produce this exchange: actions 1 and 6 select the first entry of the five-entry path through modulo indexing. The zero actions leave the configuration unchanged at the remaining stages. These two tuples form the complete alias set of the resulting configuration, illustrating how distinct register codes can represent the same switch configuration.}
\label{fig:case-1-example}
\end{figure*}

\emph{Radiality.} A pass operation $a_s=0$ preserves the current tree and progresses directly to the next stage action in $\bm a$. Closing a tie creates one cycle on the graph over the tie and its endpoint path. Removing any eligible path-edge restores a connected tree, and induction over stages therefore proves radiality of every decoded branch-exchange action. Removing the fixed super-root edges leaves one source in each component, as required by Eq.~\eqref{eq:radial_set}.

\emph{Action space and coverage.} Stage $s$ uses $w_s$ bits and therefore admits $K_s=2^{w_s}$ action values, including the pass operation $a_s=0$. The complete action space is
\begin{equation}
\mathcal A=\prod_{s=0}^{p-1}\{0,\ldots,K_s-1\},
\qquad
|\mathcal A|=\prod_{s=0}^{p-1}K_s=2^q.
\label{eq:action_space}
\end{equation}
Each action tuple $\bm a=(a_0,\ldots,a_{p-1})$ specifies one complete sequence of stage actions. The decoder maps these tuples to switch configurations, and its image $\mathcal R=\mathsf{Dec}(\mathcal A)$ is the set of distinct reachable topologies.

To establish coverage of radial topologies in Eq.~\eqref{eq:radial_set}, choose any radial configuration that contains all fixed edges. Here, we show how to reproduce this chosen configuration from the radial default. Assume that the register includes every default-open controlled tie, every controlled path branch may be opened, and the state widths satisfy Eq.\eqref{eq:widths}. Then, process the ties in stage order. If a tie is open in the chosen configuration, take the pass action. If it is closed in the chosen configuration, close it in the current configuration, creating one cycle which must contain another edge that is open in the chosen configuration; otherwise, the chosen configuration would also contain that cycle. This edge can then be opened, and the width condition ensures that this opening is represented in the register. After all stages, the current tree contains every edge of the chosen configuration. This consequently establishes coverage of all radial configurations and $\mathcal R=\mathcal F$ under these assumptions. The effects of fixed-switch restrictions and rolling-window searches on coverage are discussed separately in the sections that describe these extensions to larger networks. 

\emph{Alias sets.} Note that distinct action tuples can decode to the same configuration, as evidenced in the caption to Fig. \ref{fig:case-1-example}. For $\bm z\in\mathcal R$, define
\begin{equation}
\mathcal O(\bm z)
=\{\bm a\in\mathcal A:\mathsf{Dec}(\bm a)=\bm z\},
\qquad
d(\bm z)=|\mathcal O(\bm z)|.
\label{eq:orbit}
\end{equation}
We call $\mathcal O(\bm z)$ the complete \emph{alias set} and $d(\bm z)$ its degeneracy. These aliases are different branch-exchange representations of the same switch configuration.

\subsection{Surrogate-assisted Warm-start QITE}\label{sec:workflow}
\emph{Fit and rank configurations.}
Learning-based reconfiguration has precedent in neural-network methods with clustered training sets~\cite{1645224}. More generally, combinatorial surrogate-assisted optimization can reduce expensive objective evaluations~\cite{baptista2018bocs, kitai2020fmqa}. Here, the model estimates loss on stage actions and explicitly balances the loss data for multiple representations of each configuration. We fit a model iteratively, performing multiple rounds of optimization. Before round $r$, let $\mathcal D_r=\{(\bm z_i,L_i)\}_{i=1}^{n_r}$ contain all distinct usable labels. With a fixed scale $\Delta_L>0$ in kW, define
\begin{equation}
y_i=\frac{L_i-L_{\min,r}}{\Delta_L},\qquad
g_i=\frac{n_r e^{-\beta_{\rm e}y_i}}{\sum_{j=1}^{n_r}e^{-\beta_{\rm e}y_j}},
\label{eq:loss_weights}
\end{equation}
where $L_{\min,r}=\min_iL_i$ and $\beta_{\rm e}\geq0$ controls the emphasis on low losses. Targets and weights are dimensionless, and $\sum_ig_i=n_r$. The weights here serve to bias the fit accuracy on the low-loss samples in $\mathcal{D}_r$.

The second-order action model is 
\begin{equation}
\begin{aligned}
    f_{\bm\theta}(\bm a)
    ={}&\theta_0+
    \sum_{s=0}^{p-1}\sum_{k=0}^{K_s-1}
    \theta_{s,k}\delta_{a_s,k}\\
    &+\sum_{0\le s<t<p}
    \sum_{k=0}^{K_s-1}\sum_{\ell=0}^{K_t-1}
    \theta_{st,k,\ell}\delta_{a_s,k}\delta_{a_t,\ell}.
    \end{aligned}
    \label{eq:action_surrogate}
\end{equation}
Here, $K_s$ is the number of actions at stage $s$, and $\delta_{a_s,k},$ $\delta_{a_t,\ell}$ are Kronecker deltas. The fitted coefficients $\theta_0$, $\theta_{s,k}$, and $\theta_{st,k,\ell}$ represent the intercept, stage effects, and stage-pair effects, respectively. We fit all aliases of each evaluated configuration:
\begin{equation}
\begin{aligned}
\bm\theta^*_r=\arg\min_{\bm\theta}\Bigl[&
\sum_{i=1}^{n_r}\frac{g_i}{d(\bm z_i)}
\sum_{\bm a\in\mathcal O(\bm z_i)}(f_{\bm\theta}(\bm a)-y_i)^2
\\&+\alpha\|\bm\theta_{\neg0}\|_2^2\Bigr].
\end{aligned}
\label{eq:ridge}
\end{equation}
Here $\alpha\geq 0$ is the ridge penalty and $\bm\theta_{\neg0}$ excludes the intercept from regularization. Each configuration contributes total weight $g_i$, independent of its alias count which penalizes disagreement among aliases without forcing identical predictions. Candidates are then ranked by
\begin{equation}
\begin{aligned}
\widehat y_r(\bm z)&=\frac{1}{d(\bm z)}\sum_{\bm a\in\mathcal O(\bm z)}f_{\bm\theta^*_r}(\bm a),\\
\widehat L_r(\bm z)&=L_{\min,r}+\Delta_L\widehat y_r(\bm z).
\end{aligned}
\label{eq:orbit_mean}
\end{equation}
The average in Eq.~\eqref{eq:orbit_mean} includes the complete alias set.

\emph{Map the surrogate to a diagonal Hamiltonian.} Let $b_v(k)$ be bit $v$ of action $k$ and $Z_u$ the Pauli-$Z$ operator on qubit $u$. The projector for action $k$ at stage $s$ is
\begin{equation}
\Pi_{s,k}=\prod_{v=0}^{w_s-1}\frac{I+(-1)^{b_v(k)}Z_{o_s+v}}{2},
\label{eq:projector}
\end{equation}
where $o_s$ is the register offset for the considered action from Eq.~\eqref{eq:widths}, and
\begin{equation}
    \Pi_{s,k}|\bm x(\bm a)\rangle = \delta_{a_s,k}|\bm x(\bm a)\rangle.
\end{equation}
Replacing $\delta_{a_s,k}$ and $\delta_{a_t,\ell}$ in Eq.~\eqref{eq:action_surrogate} gives
\begin{align}
    H_r&=\left( \theta^*_{r}\right)_0I+\sum_{s,k}\left(\theta^*_{r}\right)_{s,k}\Pi_{s,k}\nonumber\\
    &\quad+\sum_{s<t}\sum_{k,l}(\theta^*_r)_{st,k,l}\Pi_{s,k}\Pi_{t,l},\label{eq:hamiltonian}\\
    H_r|\bm x(\bm a)\rangle&=f_{\bm\theta^*_r}(\bm a)|\bm x(\bm a)\rangle.
    \label{eq:energy_equivalence}
\end{align}
Eq.~\eqref{eq:energy_equivalence} is exact for the fitted action model. Expansion and aggregation yields $H_r=\sum_U h_{U,r} Z_U$, where $U$ is a qubit-index set, $Z_U=\prod_{u\in U}Z_u$, and $Z_\varnothing=I$ where all terms are mutually commutative. It is important to note that the second-order action model is higher order in regards to its binary form; a stage-pair term has a Pauli locality $w_s+w_t$. We use $H_r$ to calculate the QITE update classically rather than implement its evolution directly. The resulting circuit uses two-qubit generator rotations whose parameters are determined by first-order expectation matching, as described below.

\emph{Quantum imaginary time evolution.} QITE approximates normalized imaginary-time evolution by state-dependent unitary updates~\cite{QITE_ref}. For an initial state $|\psi_{0,r}\rangle$ and our fixed surrogate Hamiltonian $H_r$, the imaginary-time evolution is
\begin{equation}
    |\psi_\mathrm{ITE}\rangle = \frac{e^{-\tau H_r}|\psi_{0,r}\rangle}{\|e^{-\tau H_r}|\psi_{0,r}\rangle\|_2}, \quad \tau\geq 0,
    \label{eq:normalized_ite}
\end{equation}
where $\tau$ is the imaginary-time parameter and this evolution increases the probability mass on lower-energy states. We use analytic expectations of a product state to determine an update following Ref.~\cite{lotshaw2026}, from which we inherit the warm-start incumbent construction and choice of generator and supply an extension to consider expectations for a Hamiltonian with arbitrary Pauli-Z support. We perform optimization through several rounds, acquiring ACPF labels through surrogate-assisted quantum search.

\emph{Initial state.} At acquisition round $r$, we select an incumbent configuration to initialize our quantum state. Specifically, the incumbent is the configuration with the lowest measured loss in $\mathcal D_r$.  Because our quantum register represents branch-exchange actions and is generally many-to-one as established in Eq.~\eqref{eq:orbit}, we select the action tuple $\bm a$ with the lowest computed Hamiltonian energy from within the incumbent's alias set. This choice is needed because aliases of one configuration can have different fitted surrogate energies.

Let $x_u^\star$ be bit $u$ of the selected register code and prepare the product state
\begin{equation}
    \begin{aligned}
        |\chi_u\rangle&=\cos\phi_r|x_u^\star\rangle+\sin\phi_r|1-x_u^\star\rangle,\\
        |\psi_{0,r}\rangle&=|\chi_{q-1}\rangle\otimes\cdots\otimes|\chi_0\rangle.
    \end{aligned}
    \label{eq:warm_state}
\end{equation}
Starting from $|0\rangle$, qubit $u$ receives $R_y(2\phi_r)$ followed by $X_u$ when $x_u^\star=1$, where $X,\ Y,\text{and } Z$ are the Pauli operators and $R_y(\theta)=e^{-\mathrm i \theta Y/2}$. Each bit then matches the incumbent with probability $\cos^2\phi_r$ as the angle $\phi_r$ controls the spread around the selected incumbent's representative binary form. At $\phi_r=\pi/4$ the distribution is uniform over all branch-exchange actions while at $\phi_r=0$ the state is a fixed point of the imaginary-time evolution. Therefore, we utilize $0<\phi_r\leq \pi/4$.

\emph{Generator and circuit.} We apply a parameterized unitary to the initial state $|\psi_{0,r}\rangle$ to approximate normalized imaginary-time evolution under the fitted surrogate Hamiltonian $H_r$. We determine the parameters by least-squares matching of the first-order changes in selected Pauli-$Z$ expectations, detailed below. For each pair $0\leq i<j<q$ of qubits, we choose the Hermitian generator
\begin{equation}
    A_{ij} = Y_iZ_j + Z_iY_j,
    \label{eq:qite_generator}
\end{equation}
where $Y_i$ and $Z_i$ denote Pauli operators on qubit $i$, with identity on unlisted qubits. The two Pauli terms within each generator commute, so each pair rotation factors into two Pauli rotations. See Fig.~\ref{fig:qite_circuit} for a detailed construction. The coefficients are computed classically from analytic expectations of the initial product state. The QITE circuit is then the fully-connected pairwise ansatz
\begin{equation}
    U^\mathrm{QITE}(\bm \mu)=\prod_{i<j}e^{-\mathrm i \mu_{ij}A_{ij}}.
    \label{eq:qite_circuit}
\end{equation}

\emph{Parameter selection.} Fix $H=H_r$ and evaluate all expectations in the initial state $|\psi_{0,r}\rangle$. For each retained nonidentity Pauli term in $H$, let $\Lambda$ denote its nonempty set of qubit indices. The operator $Z_\Lambda=\prod_{u\in\Lambda}Z_u$ acts as Pauli-$Z$ on these qubits and as the identity on all others. We define the imaginary-time target $D_\Lambda$ as one-half of the initial rate of change of $\langle Z_\Lambda\rangle$ under normalized imaginary-time evolution. Since $H$ commutes with $Z_\Lambda$, differentiating Eq.~\eqref{eq:normalized_ite} at $\tau=0$ gives
\begin{equation}
\begin{aligned}
    D_\Lambda &=\frac12\left.
    \frac{\partial\langle Z_\Lambda\rangle_{\mathrm{ITE}}}{\partial\tau} \right|_{\tau=0}\\
    &=\langle H\rangle\langle Z_\Lambda\rangle-\langle HZ_\Lambda\rangle.
\end{aligned}
\label{eq:qite_target}
\end{equation}
Thus $D_\Lambda$ specifies the desired change under imaginary-time evolution.

The response $G_{\Lambda,ij}$ is half the initial change produced by varying the pair coefficient $\mu_{ij}$. For the circuit state $U^{\mathrm{QITE}}(\bm \mu)|\psi_{0,r}\rangle$,
\begin{equation}
\begin{aligned}
    G_{\Lambda,ij}
    &=\frac12
    \left.
    \frac{\partial\langle Z_\Lambda\rangle_{\mathrm{QITE}}}
       {\partial \mu_{ij}}
    \right|_{\bm \mu=0}\\
    &=\frac{\mathrm i}{2}
    \langle[A_{ij},Z_\Lambda]\rangle\\
    &=-|\Lambda\cap\{i,j\}|\sin(2\phi_r)
    \langle Z_{\Lambda\triangle\{i,j\}}\rangle.
\end{aligned}
\label{eq:qite_response}
\end{equation}
The intersection size counts zero, one, or two shared qubits between the Hamiltonian term and the generator term. If both generator qubits occur in $\Lambda$, both contributions are retained.

For any set of qubit indices $U$,
\begin{equation}
    \langle Z_U\rangle
    =
    (-1)^{\sum_{u\in U}x_u^\star}
    \cos^{|U|}(2\phi_r),
    \qquad
    \langle Z_\varnothing\rangle=1.
    \label{eq:expectations}
\end{equation}
Also, $\langle X_u\rangle=\sin(2\phi_r)$ and $\langle Y_u\rangle=0$. Products of diagonal Pauli terms satisfy $Z_UZ_\Lambda=Z_{U\triangle\Lambda}$. For the expansion $H_r=\sum_U h_{U,r}Z_U$, these identities give
\begin{equation}
\begin{aligned}
    \langle H_r\rangle
    &=\sum_U h_{U,r}\langle Z_U\rangle,\\
    \langle H_rZ_\Lambda\rangle
    &=\sum_U h_{U,r}\langle Z_{U\triangle\Lambda}\rangle.
\end{aligned}
\label{eq:classical_qite_moments}
\end{equation}
Every expectation on the right-hand side follows analytically from Eq.~\eqref{eq:expectations}, including those with support on more than two qubits. The target and response can therefore be calculated classically without circuit measurements.

The higher-locality terms in $H_r$ contribute to this classical calculation but are not compiled into evolution gates. The quantum circuit instead prepares the product state, applies the pair rotations generated by $A_{ij}$, and measures the register. Their parameters approximate the initial imaginary-time response within the chosen generator family. Each acquisition round starts from a fresh product state, so the analytic expectation formulas apply again.

Let $\bm{\dot\mu}$ be the parameter velocity at imaginary time zero. The desired target and response derivatives are $2\bm D$ and $2 G \bm{\dot \mu}$, respectively. Matching the first-order terms therefore requires $G\dot{\bm \mu}\simeq\bm D$. Note, the restricted generator family need not reproduce every target derivative. We use a regularized least-squares fit:
\begin{equation}
\begin{aligned}
    \dot{\bm \mu} &=\arg\min_{\bm v}\left\{
    \|G\bm v-\bm D\|_2^2+\lambda\|\bm v\|_2^2\right\},\\
    \bm \mu&=\Delta\tau\,\dot{\bm \mu}.
\end{aligned}
\label{eq:qite_solve}
\end{equation}
Here $\lambda\geq0$ penalizes large parameter derivatives, and $\Delta\tau\geq0$ is the finite imaginary-time step. We solve the system by singular-value decomposition and any nonzero fitting residual indicates a first-order expectation mismatch.

\emph{Sampling.} For each shot, we prepare Eq.~\eqref{eq:warm_state}, apply Eq.~\eqref{eq:qite_circuit} using parameter updates from Eq.~\eqref{eq:qite_solve}, and measure all register qubits in the computational basis. The acquisition procedure decodes the measured bit strings, merges sampled aliases, and selects configurations for ACPF evaluation. Each subsequent round uses a newly fitted Hamiltonian and a fresh initial state on an updated incumbent. 

\begin{figure*}[!htbp]
    \centering
    \resizebox{0.85\textwidth}{!}{%
\begingroup%
\providecommand{\QITECircuitTextFont}{\fontfamily{ptm}\selectfont}%
\definecolor{qcInk}{HTML}{292929}%
\definecolor{qcMuted}{HTML}{626262}%
\definecolor{qcPurple}{HTML}{6F3C8A}%
\definecolor{qcPurplePale}{HTML}{F3EEF8}%
\definecolor{qcGrey}{HTML}{5F5F5F}%
\definecolor{qcGreyPale}{HTML}{F0F0F0}%
\begin{tikzpicture}[
  x=1mm,y=-1mm,
  every node/.style={anchor=west,inner sep=0pt,outer sep=0pt,text=qcInk,
    font=\normalfont\QITECircuitTextFont\fontsize{10}{12}\selectfont},
  qchead/.style={font=\normalfont\QITECircuitTextFont\bfseries\fontsize{10}{12}\selectfont},
  qcwire/.style={draw=qcInk,line width=0.75pt},
  qcgate/.style={draw=qcPurple,fill=qcPurplePale,line width=0.85pt},
  qcwarm/.style={draw=qcGrey,fill=qcGreyPale,line width=0.85pt},
  qcmath/.style={anchor=center,font=\normalfont\fontsize{10.95}{13.14}\selectfont},
  qcsmall/.style={font=\normalfont\QITECircuitTextFont\fontsize{9}{10.8}\selectfont}
]
\path[use as bounding box] (0,0) rectangle (180,126);

\node[qchead] at (4,5) {(a) Three-qubit QITE circuit};
\node[anchor=east] at (176,5)
  {Incumbent representative: $|\bm{x}^\star\rangle
    =|1\rangle_2\otimes|0\rangle_1\otimes|1\rangle_0$};
\node[qchead,anchor=center,text=qcGrey] at (39,13) {Warm start};
\node[anchor=center,text=qcGrey] at (39,19) {$0<\phi_r\leq\pi/4$};
\node[qchead,anchor=center,text=qcPurple] at (105,13) {One finite QITE update};
\node[anchor=center,text=qcPurple] at (105,19)
  {$U^{\mathrm{QITE}}=U_{12}U_{02}U_{01}$};
\node[qchead,anchor=center] at (158,13) {Measure};

\foreach \u/\yy in {0/27,1/41,2/55} {
  \node[anchor=east] at (19,\yy) {$q_\u:\ |0\rangle$};
  \draw[qcwire] (21,\yy) -- (155,\yy);
  \draw[qcwarm] (24,{\yy-4}) rectangle (44,{\yy+4});
  \node[qcmath,text=qcGrey] at (34,\yy) {$R_y(2\phi_r)$};
}
\foreach \yy in {27,55} {
  \draw[qcwarm] (49,{\yy-4}) rectangle (57,{\yy+4});
  \node[qcmath,text=qcGrey] at (53,\yy) {$X$};
}
\draw[draw=qcMuted,densely dotted,line width=0.65pt] (63,21) -- (63,61);

\draw[qcgate] (71,23) rectangle (90,45);
\node[qcmath,text=qcPurple] at (80.5,34) {$U_{01}$};
\draw[qcgate] (96,23) rectangle (115,59);
\node[qcmath,text=qcPurple] at (105.5,31.5) {$U_{02}$};
\draw[qcwire,draw=qcMuted] (96,41) -- (115,41);
\node[qcmath,text=qcMuted,fill=qcPurplePale,inner xsep=2pt] at (105.5,41) {$I_1$};
\draw[qcgate] (121,37) rectangle (140,59);
\node[qcmath,text=qcPurple] at (130.5,48) {$U_{12}$};

\foreach \u/\yy in {0/27,1/41,2/55} {
  \draw[qcwire,fill=white] (154,{\yy-3.5}) rectangle (161,{\yy+3.5});
  \draw[qcwire] (155.2,{\yy+1.3})
    .. controls (155.3,{\yy-2.4}) and (159.7,{\yy-2.4}) .. (159.8,{\yy+1.3});
  \draw[qcwire] (157.5,{\yy+1.3}) -- (159.1,{\yy-2.1});
  \draw[qcwire] (161,{\yy-0.35}) -- (169,{\yy-0.35});
  \draw[qcwire] (161,{\yy+0.35}) -- (169,{\yy+0.35});
  \node at (171,\yy) {$x_\u$};
}
\node[qchead,text=qcGrey] at (4,65) {Prepared register state};
\node[anchor=center] at (90,74)
  {$|\psi_{0,r}\rangle
    =\bigl(\cos\phi_r|1\rangle+\sin\phi_r|0\rangle\bigr)_2
     \otimes\bigl(\cos\phi_r|0\rangle+\sin\phi_r|1\rangle\bigr)_1
     \otimes\bigl(\cos\phi_r|1\rangle+\sin\phi_r|0\rangle\bigr)_0$};

\node[qchead] at (4,84) {(b) One pair rotation};
\foreach \yy/\ll in {94/i,104/j} {
  \node[anchor=east] at (15,\yy) {$q_\ll$};
  \draw[qcwire] (17,\yy) -- (53,\yy);
  \draw[qcwire] (72,\yy) -- (176,\yy);
}
\draw[qcgate] (23,89.5) rectangle (49,108.5);
\node[qcmath,text=qcPurple] at (36,99) {$U_{ij}(\mu_{ij})$};
\node[qcmath] at (62,99) {$=$};
\draw[qcgate,fill=white] (76,89.5) rectangle (120,108.5);
\node[qcmath,text=qcPurple] at (98,99) {$R_{Y_iZ_j}(2\mu_{ij})$};
\draw[qcgate,fill=white] (128,89.5) rectangle (172,108.5);
\node[qcmath,text=qcPurple] at (150,99) {$R_{Z_iY_j}(2\mu_{ij})$};

\node[anchor=center] at (90,117)
  {$\begin{aligned}
    A_{ij}&=Y_iZ_j+Z_iY_j,
      &\quad U_{ij}(\mu_{ij})&=e^{-\mathrm i\mu_{ij}A_{ij}},\\[3pt]
    R_P(\theta)&=e^{-\mathrm i\theta P/2},
      & [Y_iZ_j,Z_iY_j]&=0.
    \end{aligned}$};
\end{tikzpicture}%
\endgroup%
%
    }
    \caption{Warm-start QITE circuit for a three-qubit register. (a) In this example, the incumbent representative is the state $|\bm x^\star\rangle=|101\rangle$, with basis order $q_2q_1q_0$. Each qubit receives $R_y(2\phi_r)$, followed by $X$ when its incumbent bit is one, preparing the tensor-product state $|\psi_{0,r}\rangle$. The pair-wise unitaries then act from left to right in the order $(0,1),(0,2),(1,2)$, before computational-basis measurement returns the register bits. The identity $I_1$ marks the unchanged middle qubit during $U_{02}$. (b) The two commuting Pauli terms in $A_{ij}$ give the exact factorization of $U_{ij}$ into the rotations shown, each with angle $2\mu_{ij}$. The parameters are computed classically through expectation matching under the fitted surrogate Hamiltonian.}
    \label{fig:qite_circuit}
\end{figure*}
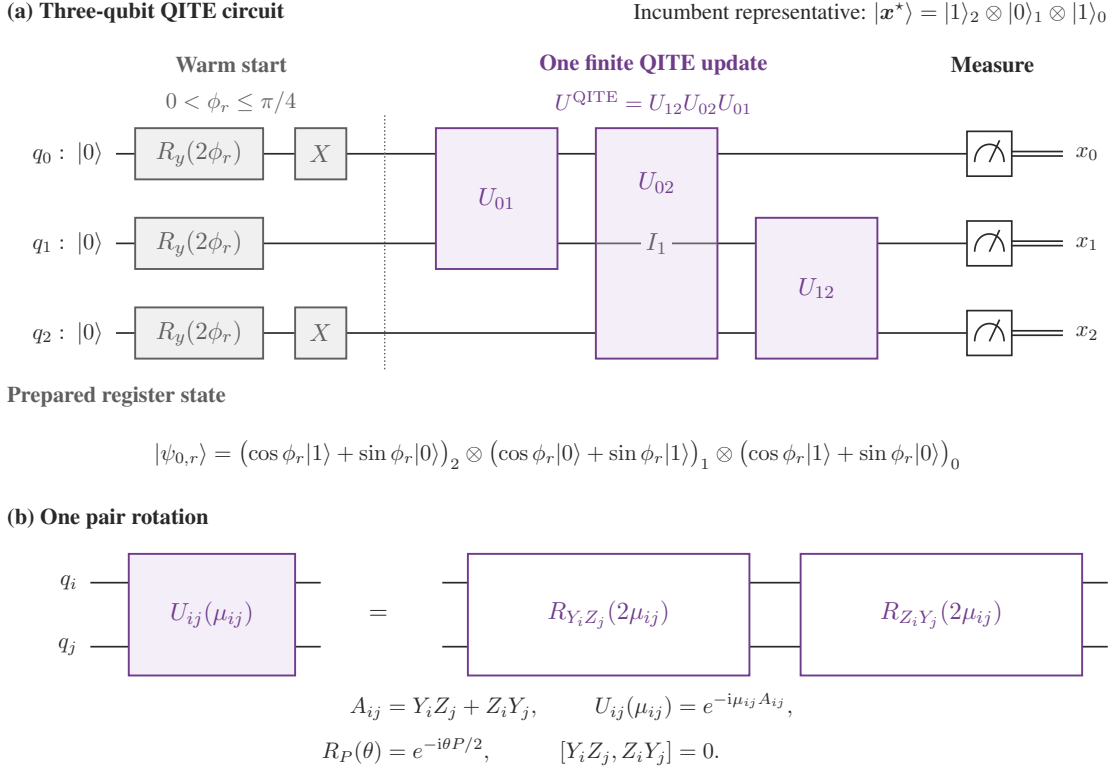

\emph{Acquire labels and update incumbent.} To initialize the surrogate fit, ACPF is solved for the default configuration and a deterministic sequence of single-stage branch-exchanges until $N_0$ usable labels are obtained or that reserve is exhausted. Let $n_{\rm init}$ be the realized count of labels at initialization. With a total label cap of $N$ and batch size $B$, set $T=\lceil(N-n_{\rm init})/B\rceil$. For $T>0$, the warm start initialization angle is annealed across rounds according to the schedule
\begin{equation}
\phi_r=\phi_{\rm start}+\frac rT(\phi_{\rm target}-\phi_{\rm start}),\quad
r=0,\ldots,T-1,
\label{eq:schedule}
\end{equation}
with $0<\phi_{\rm target}<\phi_{\rm start}\leq\pi/4$. Thus, as the rounds progress, the incumbent state is increasingly localized in  probability amplitude and the final returned state is the lowest loss ACPF state observed through all rounds.

\begin{algorithm}[t]
\caption{Surrogate-assisted QITE acquisition}\label{alg:acquisition}
\begin{algorithmic}[1]
\Require Decoder, evaluator, $N,N_0,B$, shots $S_r$, angle endpoints
\State $\mathcal D\gets\varnothing$; attempted set $\mathcal A_{\rm tried}\gets\varnothing$
\State Attempt initialization states; record all attempts and usable labels
\If{$\mathcal D=\varnothing$}\State Report initialization failure; stop\EndIf
\State Fix $T$ and angles using the realized initial-label count
\For{$r=0,\ldots,T-1$}
\If{$|\mathcal D|=N$ or an exit rule is met}\State Stop\EndIf
\State Fit Eqs.~\eqref{eq:loss_weights}--\eqref{eq:ridge}; construct $H_r$
\State Choose the measured incumbent and its minimum-energy alias
\State Prepare one QITE circuit; measure $S_r$ shots
\State Decode outcomes, merge aliases, and exclude $\mathcal A_{\rm tried}$
\If{no eligible configuration remains}\State Record pool exhaustion; stop\EndIf
\State Rank eligible topologies by Eq.~\eqref{eq:orbit_mean}
\State Attempt candidates until $\min(B,N-|\mathcal D|)$ new usable labels or pool exhaustion
\State Add every attempt to $\mathcal A_{\rm tried}$; add usable labels to $\mathcal D$
\EndFor
\State Return the lowest measured loss in $\mathcal D$ and the stop reason
\end{algorithmic}
\end{algorithm}

\subsection{Extending to Larger Networks} \label{sec:larger_extensions}
The full branch-exchange register can exceed the available qubit count for larger distribution networks. We therefore introduce two heuristic methods that reduce register size. A DISTOP-inspired switch-fixing technique~\cite{25637} holds selected branches closed, reducing the number of eligible branch-exchange actions. A rolling-window method instead optimizes successive subsets of tie stages and carries accepted configuration changes between windows.

Both methods retain the branch-exchange decoder and evaluate each candidate on the full network using classical ACPF. They reduce the quantum register size by restricting the search space, which may exclude the minimum-loss configuration of the unrestricted problem. The following passages describe how each restriction is constructed and incorporated into the optimization workflow.

Fixed-switch restrictions and subsets of active ties can reduce total coverage of the feasible set in Eq.~\eqref{eq:radial_set}. For a fixed restricted construction, let $\mathcal F_R\subseteq\mathcal F$ denote the radial topologies consistent with these restrictions; thus, $\mathcal R\subseteq\mathcal F_R$.

\emph{Reduced-register cases.} The DISTOP method introduced by Shirmohammadi and Hong \cite{25637} is a heuristic technique to solve DNR. Built upon work of Merlin and Back \cite{merlin1975}, it constructs a network state hoped to be near-optimal by starting from the meshed network, ignoring the reactance of each branch, and consecutively opening the switch with the lowest magnitude of current flow that is part of a cycle until a radial configuration is reached. Key to this method is the idea that the radial configuration with the lowest losses is as similar as possible to the meshed configuration in terms of the current flows in the system; opening switches with the lowest current disturbs the pattern of flows the least, resulting in a radial configuration with a current flow pattern close to that of the meshed configuration. This heuristic separates the switches into two classes: those with low magnitudes of current flow that are likely to participate in reconfiguration, and those with high magnitudes of current flow that are likely to remain fixed closed. Adopting this idea, once we have solved ACPF for the meshed network and ranked the switches according to their magnitude of current flow, we can summarily fix closed some proportion of the switches with the largest current flows, reducing the size of the problem. We have set the proportion of switches fixed in this manner to $25\%$, allowing a modest reduction in problem size. Importantly, this method is only reliable if none of the switches in the optimal configuration of a dataset is fixed. An analysis of the test cases we consider in this work shows that none of the switches open in their optimal configurations are found within the $25\%$ of switches we fix, and so we view our $25\%$ limit as resulting in a very effective heuristic when only a slight reduction in problem size or improvement in circuit depth is desired.

\emph{Rolling-Window method.} The available qubit count limits the number of branch-exchange stages that can be represented in one quantum register. To address larger networks, we optimize subsets of stages in sequence and carry incumbents forward between subsets. Although arbitrary subsets can be used, we adopt a rolling-window approach.

Each window selects consecutive stages according to an ordered list of tie switches. Only the selected stages enter the quantum register; the remaining open ties are held fixed open during that window. This reduces the register size while retaining the full network in each classical ACPF evaluation. Accepted improvements update the baseline configuration and the list of ties. The window then advances with a prescribed overlap, wrapping around the list as needed to complete a sweep. The window size limits the number of stages optimized simultaneously and is chosen to satisfy the qubit budget. Smaller windows generally require more subproblems and circuit executions to complete the sweep.

Algorithm ~\ref{alg:rolling} summarizes the rolling-window method. Let $p_{\mathrm{win}}$ be the maximum number of active stages and $p_{\mathrm{adv}}$ the step size for advancing the queue, with $1 \leq p_{\mathrm{adv}}\leq p_{\mathrm{win}}$. We perform at most $W$ windows, indexed by $\kappa$. 

\begin{algorithm}[t]
    \caption{Rolling-window surrogate-assisted QITE}
    \label{alg:rolling}
    \begin{algorithmic}[1]
        \Require Initial radial configuration $\bm z^{(0)}$;  ACPF evaluator, rolling-window parameters, and per window acquisition settings
        \State $\bm z_{\mathrm{default}}\gets\bm z^{(0)}$
        \State Initialize the best observed configuration as unset
        \State $\mathcal Q\gets$ ordered queue of open switches in $\bm z_{\mathrm{default}}$
        \For{$\kappa=1,\ldots,W$}
            \State Select an active set $\mathcal{W}$ of $\mathcal Q$ using the register limits
            \If{no valid window exists}
                \State Report window-selection failure; stop
            \EndIf
            \State Construct the decoder from $\bm z_{\mathrm{default}}$ and active ties $\mathcal{W}$
            \State Hold all other currently open switches open
            \State Run Algorithm~\ref{alg:acquisition}; retain usable labels $\mathcal D^{(\kappa)}$
            \If{$\bm z_\mathrm{default}$ has no usable label}
                \State Report default-evaluation failure; stop
            \EndIf
            \State $\bm z_{\mathrm{cand}}\gets$ lowest-loss configuration in $\mathcal{D}^{(\kappa)}$
            \State Update the best observed configuration $\bm z_{\mathrm{best}}$ using $\bm z_{\mathrm{cand}}$
            \State $\bm z_{\mathrm{next}}\gets\bm z_{\mathrm{cand}}$
            \State $\mathcal W_{\mathrm{adv}}\gets$ first $\min(p_{\mathrm{adv}},|\mathcal W|)$ entries of $\mathcal Q$
            \State Remove from $\mathcal Q$ switches closed in $\bm z_{\mathrm{next}}$
            \State Move remaining entries of $\mathcal W_{\mathrm{adv}}$ to the end of $\mathcal{Q}$
            \State Append switches closed in $\bm z_{\mathrm{default}}$ and open in $\bm z_{\mathrm{next}}$, in switch-vector order
            \State $\bm z_{\mathrm{default}}\gets \bm z_{\mathrm{next}}$
        \EndFor
        \State Return $\bm z_{\mathrm{best}}$ and its measured loss
    \end{algorithmic}
\end{algorithm}

\section{Results}\label{sec:results}
In this work, ideal simulations used local Qiskit Aer \cite{qiskit2024} with finite-shot sampling and no device-noise model. Noisy simulations used the IonQ cloud simulator \cite{ionq_backends} through the Qiskit-IonQ interface with either $\texttt{forte-1}$ or $\texttt{forte-enterprise-1}$ noise model matched to the hardware target. Hardware runs used the IonQ $\texttt{forte-1}$ or $\texttt{forte-enterprise-1}$ backends \cite{chen2024forte}. Both are trapped-ion processors with all-to-all qubit connectivity. Computational-basis states were sampled from the quantum circuit while surrogate fitting, QITE parameter calculation, branch-exchange decoding, and ACPF evaluation were conducted classically. For the considered cases, 100 ideal simulations, 25 noisy simulations, and 1 hardware run were performed. Noisy simulation and QPU results are presented in the main text while treatment for the ideal simulations is deferred to Appendix~\ref{app:ideal_sims}.

\subsection{Test Cases 1-5}
We begin with presentation of Test Cases 1-5: Case 1 is a 14 bus case adapted from the standard IEEE 16 bus case \cite{193906} through consolidation of its source nodes, Cases 2 and 4 are the standard IEEE 33 and 69 bus cases originated by Baran and Wu \cite{25627, 19265}, and Cases 3 and 5 are the reductions of the 33 and 69 bus cases according to the register-reduction method of Section \ref{sec:larger_extensions}. We apply the standard algorithm described in Fig. \ref{fig:workflow} to these cases, without employing the rolling-window approach. Reducing the size of the branch-exchange encoding for the larger two networks allows us to decrease the width and depth of the circuits used in their optimization, reducing quantum circuit error. The experimental setup for each dataset is presented in Table \ref{tab:standard_case_settings}. We exhibit reductions in the total number of ACPF solves and shots required to reach the optimal configurations for each reduced dataset compared to their base cases. 

Across Cases 1--5, the workflow reduces losses from the default configurations and identifies an optimal configuration in every hardware run. Table~\ref{tab:standard_case_results} summarizes the default and minimal-loss configurations, their losses, and the hardware-run evaluation counts required for first discovery. Cases 2 and 3 share the same minimal-loss configurations, as do Cases 4 and 5, though the minimal-loss configuration discovered in the QPU run differs between Cases 4 and 5. The hardware runs on the reduced-register cases 3 and 5 reach these minimal-loss configurations with fewer ACPF evaluations. However, this improvement is not uniform across the noisy simulation ensembles. These results demonstrate the operation of the complete workflow within the selected evaluation budgets, while indicating that the benefits of reducing circuit resources depend on the restricted search space. 

\begin{widetable}
\centering
\caption{Network sizes, register allocations, and acquisition budgets
for Cases 1--5. Stage widths give the number of qubits assigned
to each branch-exchange stage.}
\label{tab:standard_case_settings}
\small
\setlength{\tabcolsep}{4pt}
\renewcommand{\arraystretch}{1.15}

\begin{tabular*}{\textwidth}{
    @{\extracolsep{\fill}} c c c c c c c c c c @{}
}
\toprule
& \multicolumn{3}{c}{Network}
& \multicolumn{2}{c}{Register}
& \multicolumn{4}{c}{Acquisition settings} \\
\cmidrule(lr){2-4}
\cmidrule(lr){5-6}
\cmidrule(l){7-10}

Case
& Buses
& Switches
& Ties
& Qubits
& Stage widths
& Circuits
& \shortstack{Initial ACPF\\budget}
& \shortstack{ACPF budget\\per circuit}
& \shortstack{Shots per\\circuit} \\
\midrule

1 & 14 & 16 & 3 & 11 & $(3,4,4)$
  & 19 & 1  & 1                       & 10    \\

2 & 33 & 37 & 5 & 21 & $(4,3,4,5,5)$
  & 10 & 60 & 20\textsuperscript{a}    & 2,000 \\

3 & 33 & 27 & 5 & 19 & $(2,3,4,5,5)$
  & 10 & 60 & 20\textsuperscript{a}    & 2,000 \\

4 & 69 & 73 & 5 & 26 & $(5,4,5,6,6)$
  & 10 & 60 & 44                      & 2,000 \\

5 & 69 & 54 & 5 & 21 & $(3,4,5,3,6)$
  & 10 & 60 & 44                      & 2,000 \\

\bottomrule
\end{tabular*}

\smallskip
\parbox{\textwidth}{\footnotesize
\textsuperscript{a}The ACPF budget is 20 evaluations for each of
the first nine circuits and 10 evaluations for the final circuit.}
\end{widetable}

\begin{widefigure}
    \centering
    \includegraphics[width=\textwidth]{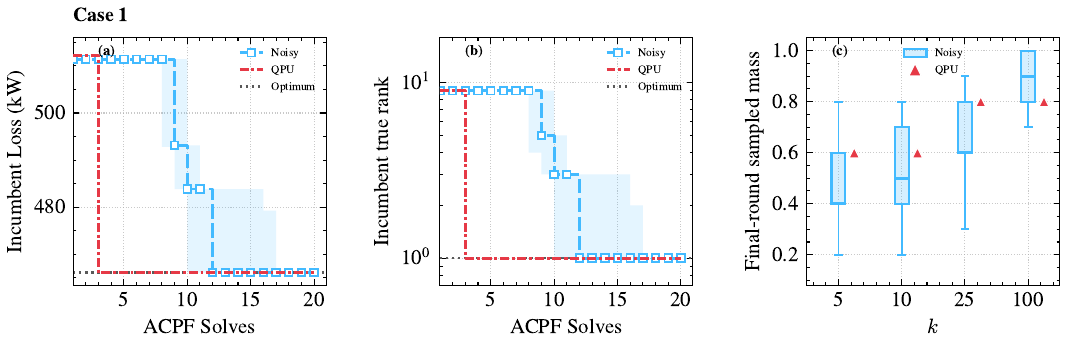}
    \caption{Results for Case 1. Panels (a) and (b) show incumbent loss and overall rank as configurations are evaluated, with the noisy simulation median and interquartile range (IQR) shown in blue and the single hardware trajectory in red. Dotted horizontal lines indicate the optimum. Panel (c) shows the fraction of final-round samples that decode into the $k$ lowest-loss configurations; boxes summarize the noisy simulation distribution, and triangles indicate the hardware result.}
    \label{fig:case1_results}
\end{widefigure}

\subsubsection{Case 1}

Case 1, shown in Fig.~\ref{fig:case-1-example}, requires an 11-qubit register and has 190 distinct radial configurations. Starting from the radial default configuration, each acquisition round uses ten circuit samples to select one additional configuration for ACPF evaluation, giving a total budget of twenty evaluated configurations under the settings in Tab.~\ref{tab:standard_case_settings}.

Within this budget, the optimal configuration with switches $7,\ 8,\text{ and } 16$ open, was identified in 22 of the 25 noisy simulation runs and in the single QPU run. As shown in Fig.\ref{fig:case1_results}(a,b), the hardware trajectory reaches this configuration at the third ACPF evaluation, including the initial evaluation of the default configuration, whereas the noisy simulation median reaches rank one at the twelfth evaluation. Additional hardware runs would be needed to determine whether this improvement persists across repeated trials.

The final-round sampling distributions in Fig.~\ref{fig:case1_results}(c) complement these histories by showing the fraction of samples from the final circuit that decode into low-loss configurations. In the hardware run, six of the ten samples belong to the five lowest-loss configurations, while eight belong to the twenty-five with lowest losses. Expanding the reference set to the one hundred configurations with lowest losses does not increase this fraction. This metric is a measure of the effectiveness of the algorithm in concentrating probability mass on states decoding to configurations with low line losses.

\subsubsection{Cases 2 \& 3}
Cases 2 and 3 are each 33-bus networks as shown in Fig.~\ref{fig:ieee33}, and differ only in which lines are switchable. For Case 2 all 37 lines are switchable, whereas Case 3 fixes ten lines closed, reducing the width of the register from 21 to 19 qubits.

As shown in Fig.~\ref{fig:case2_3_results}, both QPU runs recover the optimum, requiring 204 distinct evaluated configurations in Case 2 and 83 in Case 3, while the noisy simulation median rank traces reach rank one after 192 and 85 evaluations, respectively. The reduced Cases therefore reach the optimum earlier in their trajectories for both the hardware and the simulation. The hardware sampled mass within the best five configurations is $19.2\%$ for Case 2 and $12.5\%$ for Case 3, increasing to $24.2\%$ and $20.3\%$ within the best 100 of their respective catalogs. These results distinguish cumulative search progress from the concentration of the final sampling distribution. 

\begin{figure*}[!p]
    \centering
    \includegraphics[width=\textwidth]{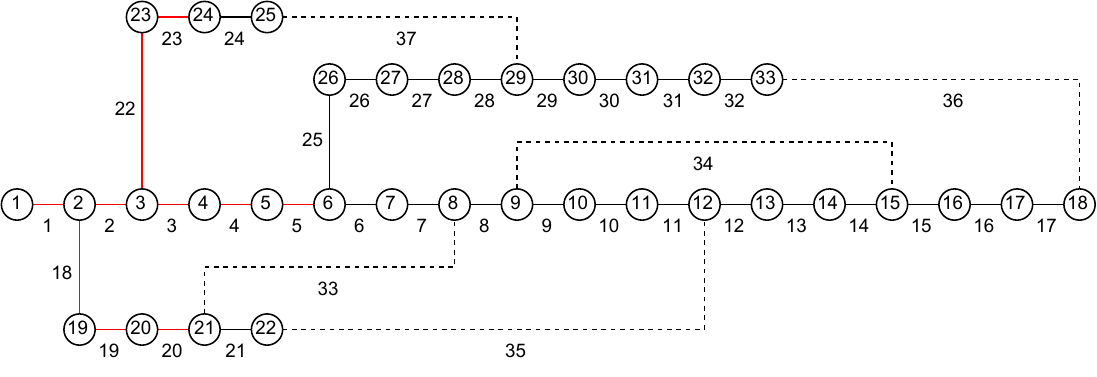}
    \caption{Cases 2 and 3 are based off the same network, shown above. In Case 2 all lines are switchable, while for Case 3 the lines marked red are fixed closed.}
    \label{fig:ieee33}

\vspace{10pt}

    \centering
    \includegraphics[width=\textwidth]{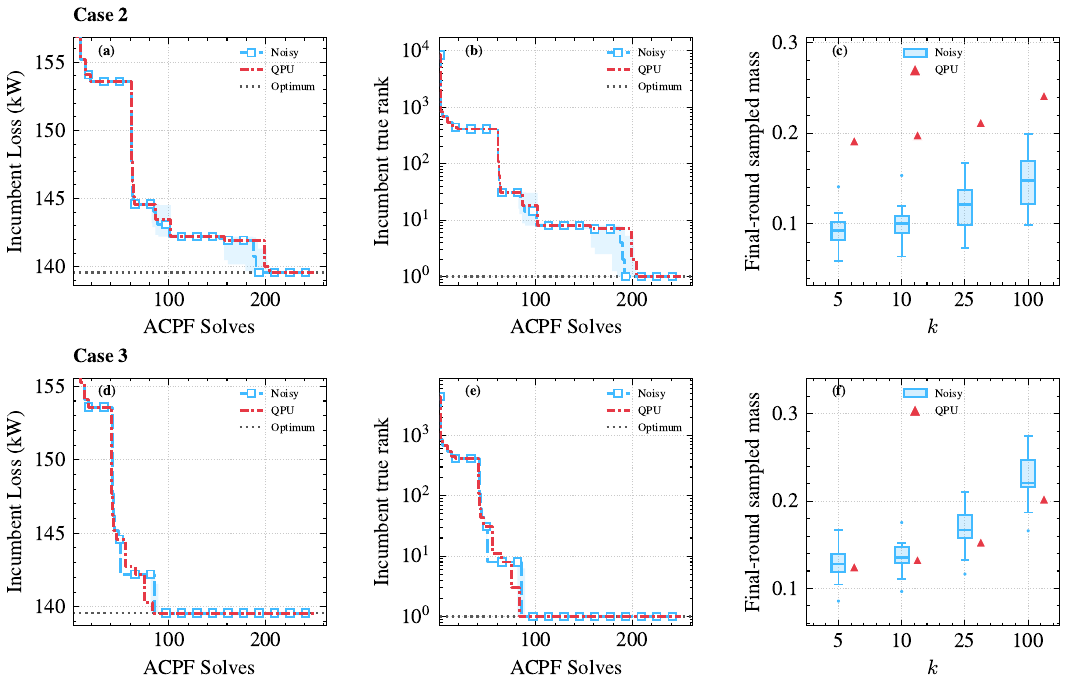}
    \caption{Search results for Case 2 (top row) and Case 3 (bottom row). Panels (a,d) show the best observed active loss, and panels (b,e) show its rank as a function of the number of distinct evaluated configurations. Noisy simulation curves show the median with IQR shading; each hardware curve represents one run, and dotted lines mark the optimum. Panels (c,f) summarize the final-round fraction of samples that decode to the best $k$ configurations in each catalog, using box plots for noisy simulations and triangles for hardware. The reduced case reaches the reference optimum earlier in the reported history, despite lower hardware sampled mass in these final-round sets.}
    \label{fig:case2_3_results}
\end{figure*}

\subsubsection{Cases 4 \& 5}
Cases 4 and 5 are each 69-bus networks as shown in Fig.~\ref{fig:ieee69}. Fixing 19 branches closed in Case 5 reduces the register width from 26 to 21 while retaining all five tie switches. Both search spaces contain the same four distinct configurations whose losses coincide at the same minimum to numerical precision. These configurations therefore share rank one.

As shown in Fig.\ref{fig:case4_5_results}, hardware reaches the minimum-loss group after 369 evaluated configurations in Case 4 and 173 in Case 5, whereas the noisy simulation median rank histories reach rank one after 281 and 380 evaluations, respectively. The reduced case also achieves greater final-round hardware sampled mass: at $k=5$ and $k=100$, the fractions increase from $0.10\%$ and $0.25\%$ to $6.7\%$ and $10.1\%$. Circuit depth is a practical obstacle for the fully connected ansatz, motivating register reduction; however, these single hardware histories do not isolate the effect of depth or establish a general improvement in search performance. 

\begin{figure*}[!p]
    \centering
    \includegraphics[width=\textwidth]{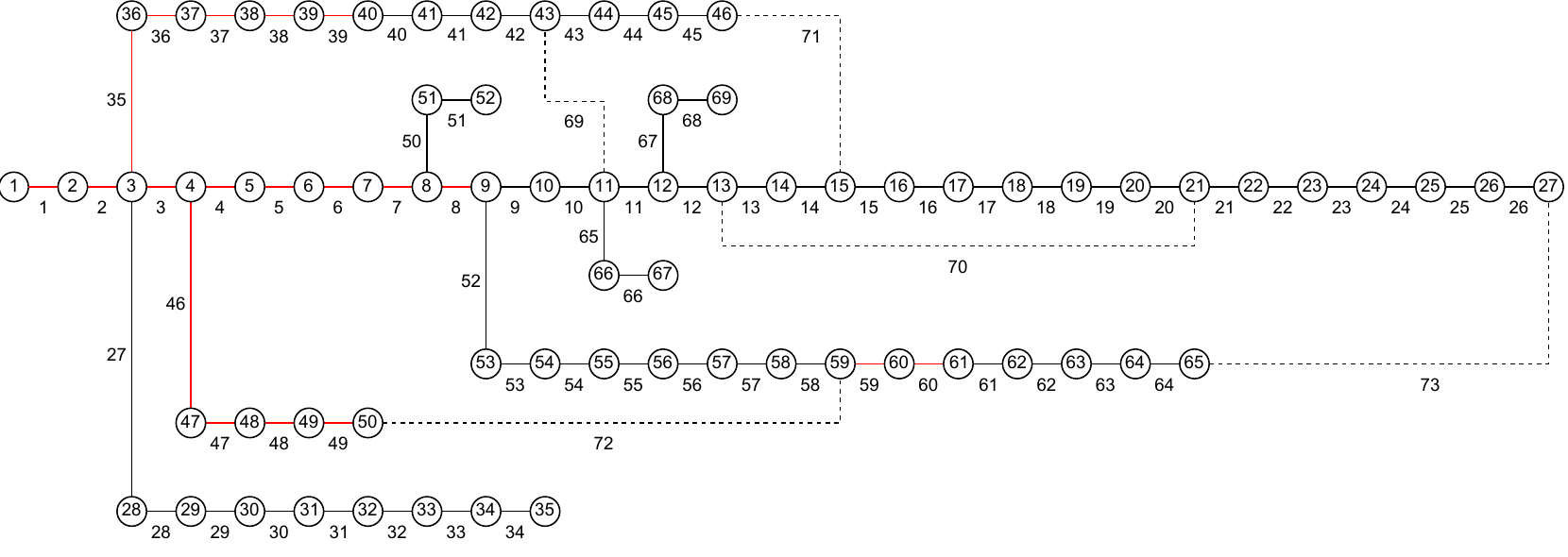}
    \caption{The network used for Cases 4 and 5. All lines are switchable for Case 4, but the red lines are fixed closed for Case 5. Note that bridges can never be opened by the branch-exchange procedure, so additionally lines 27-34, 50, 51, and 65-68 are effectively fixed closed.}
    \label{fig:ieee69}

\vspace{10pt}

    \centering
    \includegraphics[width=\textwidth]{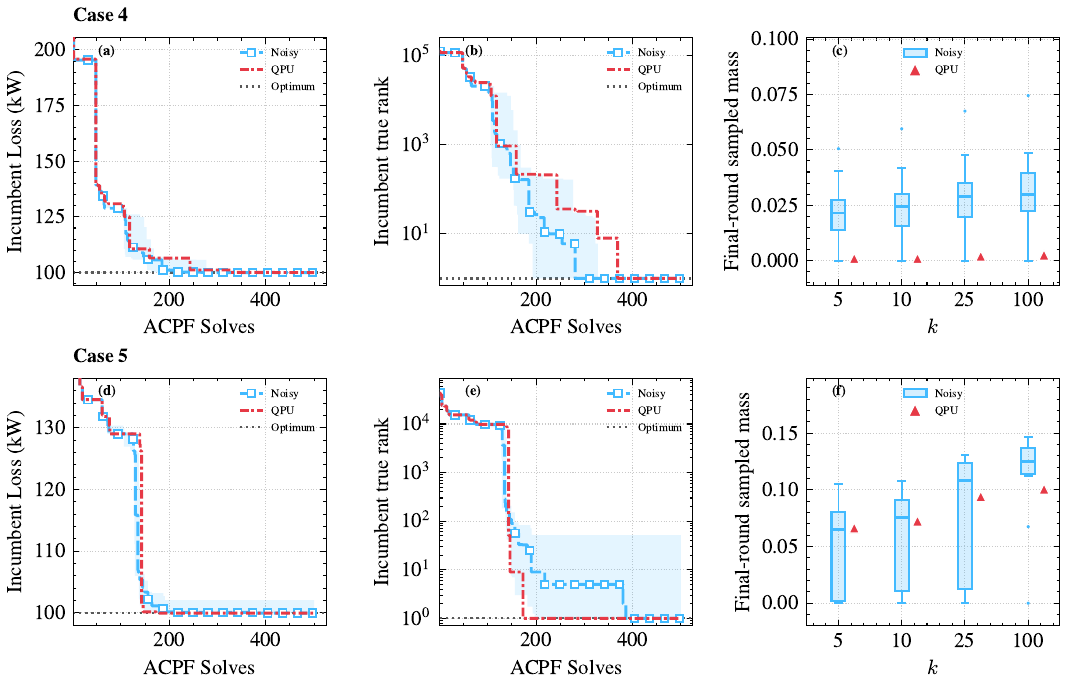}
    \caption{Results for Case 4 (top row) and Case 5 (bottom row), showing incumbent loss, rank, and final-round sampled mass. Simulation curves show medians and IQRs; hardware curves represent one run per case. Box plots summarize simulated sampled masses, with triangles marking hardware results. Equal-loss configurations share a rank, and such groupings are reflected in the top-$k$ sets.}
    \label{fig:case4_5_results}
\end{figure*}

\begin{widetable}
\centering
\caption{Radial defaults and minimal-loss configurations for Cases 1--5, specified by open-switch indices, with active losses in kW. The final column reports the number of distinct configurations evaluated with ACPF at the point each QPU run first reaches a minimal-loss configuration. Dashed lines indicate equivalence to previous entries.}
\label{tab:standard_case_results}
\small
\setlength{\tabcolsep}{5pt}
\renewcommand{\arraystretch}{1.15}
\begin{tabular*}{\textwidth}{
    @{\extracolsep{\fill}} c l l c c c @{}
}
\toprule
& \multicolumn{2}{c}{Open switches}
& \multicolumn{2}{c}{Loss (kW)}
& \\
\cmidrule(lr){2-3}
\cmidrule(lr){4-5}
Case
& Default
& Optimal
& Default
& Optimal
& \shortstack{ACPF evaluations\\to optimal} \\
\midrule
1 & 14, 15, 16
  & 7, 8, 16
  & 511.44 & 466.13 & 3 \\
2 & 33, 34, 35, 36, 37
  & 7, 9, 14, 32, 37
  & 202.68 & 139.55 & 204 \\
3 & \textemdash
  & \textemdash
  & \textemdash & \textemdash & 83 \\
4 & 69, 70, 71, 72, 73 
  & 14, 57, 61, 69, 70
  & 224.95 & 99.60 & 369 \\
5 & \textemdash
  & 14, 55, 61, 69, 70
  & \textemdash & 99.60 & 173 \\
\bottomrule
\end{tabular*}
\end{widetable}

\subsection{Test Cases 6-10}
Test Cases 6-10 are in order the 70 \cite{1564201}, 84 \cite{1208393}, 119 \cite{ZHANG2007685}, 202 \cite{CASTRO1985155}, and 415 \cite{667402} bus cases. Following the rolling-window procedure outlined in Section \ref{sec:larger_extensions}, we fix all tie switches open except those corresponding to a contiguous window of the stages and sequentially optimize as we move the window through the list of stages. We use fewer circuits for any given window than we typically use for the smaller cases to partially compensate for the larger number of optimization passes required. We also take 2000 shots per circuit for these data sets, with the rest of their experimental parameters given in Table \ref{tab:case_summary_2}.

\begin{widetable}
\centering
\caption{Network sizes and rolling-window settings for Cases 6--10.
Overlap is measured in shared ties between consecutive windows.
The ACPF budgets specify the initial evaluations per window and
the additional evaluations per circuit.}
\label{tab:case_summary_2}
\small
\setlength{\tabcolsep}{4pt}
\renewcommand{\arraystretch}{1.15}

\begin{tabular*}{\textwidth}{
    @{\extracolsep{\fill}} c c c c c c c c c c @{}
}
\toprule
& \multicolumn{3}{c}{Network}
& \multicolumn{4}{c}{Rolling-window settings}
& \multicolumn{2}{c}{ACPF budgets} \\
\cmidrule(lr){2-4}
\cmidrule(lr){5-8}
\cmidrule(l){9-10}

Case
& Buses
& Switches
& Ties
& Windows
& \shortstack{Ties per\\window}
& \shortstack{Overlap\\(ties)}
& \shortstack{Circuits per\\window}
& \shortstack{Initial per\\window}
& \shortstack{Additional per\\circuit} \\
\midrule

6  & 70  & 79  & 11 & 5  & 4 & 2 & 6 & 40 & 20 \\
7  & 84  & 96  & 13 & 4  & 4 & 0 & 3 & 20 & 20 \\
8  & 119 & 133 & 15 & 11 & 5 & 4 & 3 & 60 & 20 \\
9  & 202 & 216 & 15 & 10 & 5 & 3 & 4 & 40 & 40 \\
10 & 415 & 473 & 59 & 15 & 4 & 0 & 2 & 40 & 30 \\

\bottomrule
\end{tabular*}
\end{widetable}

\begin{widefigure}
\includegraphics[width=0.48\textwidth]{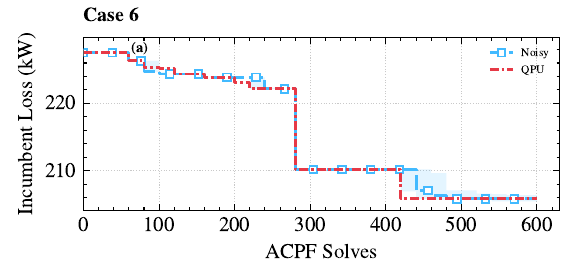}\hfill
\includegraphics[width=0.48\textwidth]{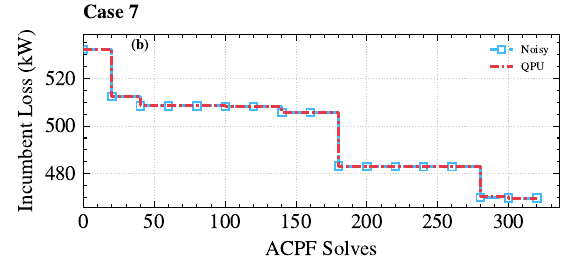}\par\vspace{4pt}
\includegraphics[width=0.48\textwidth]{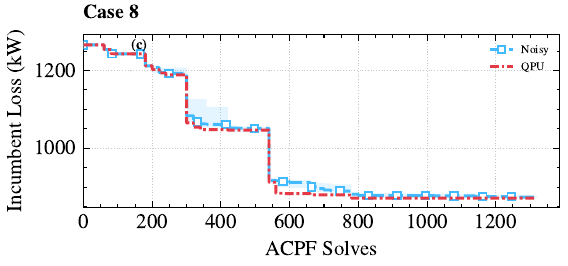}\hfill
\includegraphics[width=0.48\textwidth]{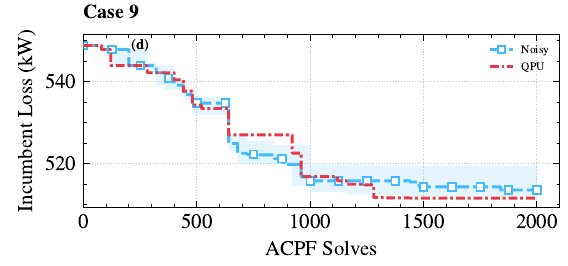}\par\vspace{4pt}
\makebox[\textwidth][l]{\includegraphics[width=0.48\textwidth]{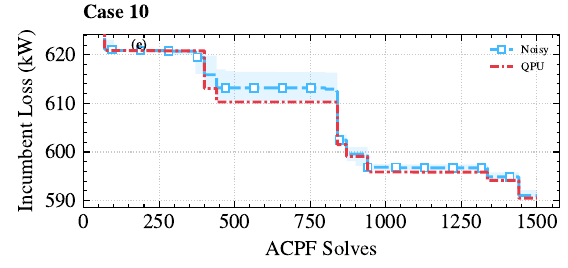}}
\caption{Best observed line loss (incumbent loss) versus cumulative ACPF solves for rolling-window Cases 6--10, shown in panels (a)--(e), respectively. Each case is optimized by moving a window of active tie-switch stages through the ordered register, as described in Sec.~\ref{sec:larger_extensions}. Blue curves show noisy simulations, while red curves show one QPU trajectory per case. For the simulation results, the curves and shaded bands give the median and interquartile range across random seeds.}
    \label{fig:rolling_traces}

\end{widefigure}

Figure~\ref{fig:rolling_traces} shows reductions in loss across all five cases for which the rolling-window approach was used, with periods of little change followed by further improvements as the search proceeds. The simulation and hardware histories follow similar trends in Cases 7, 8, and 10, with similar traces for both noisy simulation and hardware. 

The final configurations reported in Table~\ref{tab:case_summary_4} reduce loss relative to the radial default by $6.8$--$32.9\%$, with the largest reduction occurring in Case 8. Case 7 reproduces both the published reference configuration and its reported loss, while Cases 8 and 9 reach configurations with losses within $0.2\%$ of their respective reference values. The corresponding differences for Cases 6 and 10 are $2.18\%$ and $1.29\%$. Together, these results demonstrate that successive searches over limited subsets of stages can identify lower-loss configurations on larger networks while bounding the register size. Although the literature values establish global optima, they do not provide the local optima within the restricted space of states that is accessible to the rolling-window method. 

\begin{widetable}
\centering
\caption{Switch configurations and active line losses for Cases 6--10. Configurations list the open-switch indices; ranges include both endpoints. Reference configurations and losses are taken from Ref.~\cite{9565903}. Dashed lines indicate equivalence to previous entries.}
\label{tab:case_summary_4}
\small
\setlength{\tabcolsep}{5pt}
\renewcommand{\arraystretch}{1.12}

\begin{tabularx}{\textwidth}{
    @{} r l >{\raggedright\arraybackslash}X r @{}
}
\toprule
Case & Configuration & Open switches & Loss (kW) \\
\midrule

6 & Default
  & 69--79
  & 227.53 \\
  & Reference
  & 13, 30, 45, 51, 66, 70, 75, 76, 77, 78, 79
  & 201.41 \\
  & Final
  & 14, 28, 46, 51, 64, 67, 70, 72, 75, 76, 78
  & 205.80 \\
\addlinespace[5pt]

7 & Default
  & 84--96
  & 532.00 \\
  & Reference
  & 7, 13, 34, 39, 42, 55, 62, 72, 83, 86, 89, 90, 92
  & 469.88 \\
  & Final
  & \textemdash
  & 469.88 \\
\addlinespace[5pt]

8 & Default
  & 119--133
  & 1298.09 \\
  & Reference
  & 24, 27, 35, 40, 43, 52, 59, 72, 75, 96, 98, 110,
    123, 130, 131
  & 869.73 \\
  & Final
  & 24, 26, 35, 40, 43, 51, 62, 72, 74, 77, 83, 110, 122, 126, 131
  & 871.36 \\
\addlinespace[5pt]

9 & Default
  & 202--216
  & 548.90 \\
  & Reference
  & 12, 26, 43, 82, 118, 131, 133, 140, 168, 202, 203,
    208, 212, 213, 214
  & 511.18 \\
  & Final
  & 12, 29, 66, 74, 83, 111, 118, 125, 131, 135, 136, 199, 202, 208, 211
  & 511.55 \\
\addlinespace[5pt]

10 & Default
   & 415--473
   & 708.76 \\
   & Reference
   & 1, 2, 13, 15, 16, 30, 31, 40, 41, 50, 59, 73, 75,
     82, 94, 96, 97, 111, 115, 136, 142, 150, 155, 156,
     158, 163, 168, 169, 178, 179, 191, 195, 209, 214,
     230, 254, 256, 270, 294, 314, 317, 325, 358, 362,
     385, 389, 392, 395, 403, 404, 423, 424, 426, 436,
     437, 439, 446, 449, 466
   & 583.00 \\
   & Final
   & 3, 43, 50, 51, 64, 81, 131, 136, 141, 165, 179, 197, 231, 277, 289, 345, 354, 378, 383, 415, 417, 418, 419, 420, 422, 425, 426, 428, 430, 431, 432, 433, 435, 436, 437, 438, 439, 440, 441, 442, 443, 444, 445, 446, 449, 451, 453, 456, 457, 458, 460, 462, 464, 466, 467, 468, 470, 472, 473
   & 590.50 \\

\bottomrule
\end{tabularx}
\end{widetable}

\section{Conclusion}\label{sec:conclusion}
We have introduced the branch-exchange encoding as an alternate method to encode network switch configurations as binary strings; we contrast it with the default approach using switch vectors,
where techniques to address the radiality constraint have required additional binary variables or penalty terms. We view the branch-exchange encoding as an improvement due to its inherent ability to observe the radiality constraint without these drawbacks.

We produce a quadratic surrogate with respect to the stages, resulting in terms of order up to the sum of the two largest stage widths in the branch-exchange encoding, however we only use this surrogate to classically compute parameters for a parametrized quantum circuit, relieving us from needing to implement our surrogate on hardware as a HUBO. This approach therefore allows greater tractability of the problem and reduces the impact of noise on our circuit results as every branch-exchange state represents a feasible radial network configuration.

The reduction technique introduced in Section \ref{sec:larger_extensions} is a heuristic method that allows us to generate versions of test cases with fewer variables by fixing a tunably-sized set of switches that are unlikely to participate in the reconfiguration. As the proportion of fixed switches relative to the number of total switches grows, the likelihood of fixing a switch that should be open in the optimal configuration increases. For the cases we consider, fixing $25\%$ of the switches allows us to reduce the size of the branch-exchange encoding appreciably and improve performance without impairing its ability to represent the optimal network states. Our algorithm was able to discover optimal configurations for each of the Test Cases 1--5 in simulations and on quantum computing hardware, including Cases 3 and 5 for which the reduction technique was employed.

The rolling window approach developed in Section \ref{sec:larger_extensions} defines a scheme to improve upon a base network configuration by successively optimizing adjustably-sized segments of the network, allowing our algorithm to be employed for much larger systems. By modifying the size of the network segments under consideration, we can tune the width and depth of the circuits used in our algorithm to fit within hardware constraints or noise limitations. As we carry out multiple rounds of optimization in this approach, the number of circuits we need to execute is higher compared to the number used in the default approach, and we are able to explore only a fraction of the feasible states. Though we are limited in our ability to discover the optimal configuration with this method, we have seen good performance over the larger test cases we have explored, including Case 7 where the optimal configuration was discovered. For Cases 6, 8, 9, and 10, the algorithm returned configurations with losses from $0.2\%-2.18\%$ above their respective reference values.

Though current methods of solving power flow with quantum algorithms are promising, we have found that the necessary increase in the number of variables required compared to our approach of a black-box classical solver restricts the class of systems we would be able to analyze and introduces complexity in the extraction of the line losses from the power flow solution that isn't present in our method. Given these limitations, we have demonstrated a hybrid quantum-classical workflow where ACPF is solved classically, with its outputs informing the model used in the combinatorial optimization carried out on quantum computing hardware. We do not seek to directly compare this method to either fully classical or fully quantum methods; we report ACPF solution and circuit shot budgets as a way to understand the costs of our algorithm, not to advance any claims of quantum advantage for the problem sizes we have considered, and we do not make any claims about our algorithm's performance on problem sizes beyond those we have analyzed.

We now turn to lines of research that we would like to advance in future work. We intend to test ans{\"a}tze generated by operators other than that in Eq. \ref{eq:qite_generator} and experiment with including multiple generators. Furthermore, we aim to explore extensions of the surrogate model to orders higher than quadratic with respect to the stages. Coupled with changes to the generator, this may bolster the entanglement present among the qubits of the register while also necessitating more classical computation for QITE parameter updates and deeper circuits to accommodate the larger supports of Pauli strings. Beyond increasing problem size, we aim to add complexity through the consideration of additional constraints such as branch current and thermal limits as well as the effects of regulators, capacitors, and other network structures. Additionally, we plan to explore multi-objective optimization of DNR and comparisons to established solvers for problems of this type in following work.

\section*{Acknowledgments}
This report was prepared by EPB Quantum using Federal funds under award GONANB24D219-1 from the National Institute of Standards and Technology (NIST), U.S. Department of Commerce. The statements, findings, conclusions, and recommendations are those of the authors. This material is based upon work supported by the U.S. Department
of Energy, Office of Science, Office of Advanced Scientific Computing
Research under Award Number 89243024SSC000129 and under field work
proposal ERKJ445. Additionally, we would like to thank the following people: Nowar Alashkar and Cade Kennedy of EPB Quantum for their helpful insights toward developing the theoretical framework, David Nordy of EPB for aiding our understanding of power flow, Denny Dahl of IonQ for clarifying details relevant to quantum optimization, and Daniel Claudino and Teja Kuruganti of ORNL for assistance in early development of the problem formulation.

\let\qiteSelectLanguage\selectlanguage
\renewcommand{\selectlanguage}[1]{}
\bibliography{bibliography}
\let\selectlanguage\qiteSelectLanguage

\appendix
\section{Warm-Start QITE Derivation}
\label{app:qite}

This appendix derives the finite QITE update used in each acquisition
round. Starting from a fitted diagonal surrogate Hamiltonian and the incumbent selected by ACPF loss, we prepare a product state whose spread is controlled by the warm-start angle. Following the expectation-matching approach of Ref.~\cite{lotshaw2026}, we evaluate the initial imaginary-time target and unitary response in this state, then solve a regularized linear system for the circuit parameters. The Hamiltonian and initial state remain fixed during the parameter calculation and are constructed again from the available labels in the next round.

\subsection{Incumbent and representative action}

At round $r$, the nonempty usable-label set $\mathcal D_r=\{(\bm z_i,L_i)\}_{i=1}^{n_r}$ contains distinct switch configurations $\bm z_i$ and their ACPF losses $L_i=L(\bm z_i)$ in kW. A label is usable when the solve converges and returns a finite loss. We select the configuration with the lowest usable loss as the incumbent $\bm z_r^\star$, resolving equal losses by dictionary order of the switch-vector.

Because the decoder $\mathsf{Dec}$ can map several action tuples $\bm a\in\mathcal A$ to the same configuration, we first identify the incumbent's complete alias set. For any configuration $\bm z$, this set is
\begin{equation}
 \mathcal O(\bm z)=
 \{\bm a\in\mathcal A:\mathsf{Dec}(\bm a)=\bm z\}.
 \label{eq:qite_appendix_alias_set}
\end{equation}
Each action tuple has a binary register representation $\bm x(\bm a)$. The fitted action surrogate assigns this representation the energy
\begin{equation}
 H_r|\bm x(\bm a)\rangle
 =f_{\widehat{\bm\theta}_r}(\bm a)|\bm x(\bm a)\rangle,
 \label{eq:qite_appendix_raw_energy}
\end{equation}
where $\widehat{\bm\theta}_r$ is the fitted coefficient vector and the eigenvalue is a dimensionless scaled loss prediction. Since aliases of one configuration can have different surrogate energies, we choose a minimum-energy alias as the incumbent's representative:
\begin{equation}
 \bm a_r^\star\in
 \underset{\bm a\in\mathcal O(\bm z_r^\star)}{\operatorname{arg\,min}}
 f_{\widehat{\bm\theta}_r}(\bm a).
 \label{eq:qite_appendix_raw_incumbent}
\end{equation}
The selected bits and basis state are
\begin{equation}
 \bm x^\star=\bm x(\bm a_r^\star)\in\{0,1\}^q,
 \qquad |\bm x^\star\rangle
 =|x_{q-1}^\star\rangle\otimes\cdots\otimes|x_0^\star\rangle.
 \label{eq:qite_appendix_bits}
\end{equation}
This representative specifies the register state around which the warm start is prepared; the incumbent remains defined by its evaluated ACPF loss.

\subsection{Warm-start preparation and analytic expectations}

To sample states around the selected representative, choose $0<\phi_r\leq\pi/4$ and prepare the normalized product state
\begin{equation}
 \begin{aligned}
 |\chi_u\rangle
 &=\cos\phi_r|x_u^\star\rangle
   +\sin\phi_r|1-x_u^\star\rangle,\\
 |\psi_{0,r}\rangle
 &=|\chi_{q-1}\rangle\otimes\cdots\otimes|\chi_0\rangle.
 \end{aligned}
 \label{eq:qite_appendix_warm_state}
\end{equation}
With $X,Y,Z$ denoting the Pauli operators and $R_y(\theta)=e^{-\mathrm i\theta Y/2}$, each factor is obtained from $|0\rangle$ through
\begin{equation}
 |\chi_u\rangle=X_u^{x_u^\star}R_{y,u}(2\phi_r)|0\rangle.
 \label{eq:qite_appendix_preparation}
\end{equation}
Applying the rotation before the conditional $X_u$ gives the positive real amplitudes in Eq.~\eqref{eq:qite_appendix_warm_state}, so each bit matches the representative with probability $\cos^2\phi_r$. Because the qubits are initially independent, a register string $\bm x$ has probability
\begin{equation}
 |\langle\bm x|\psi_{0,r}\rangle|^2
 =(\cos^2\phi_r)^{q-d_H(\bm x,\bm x^\star)}
  (\sin^2\phi_r)^{d_H(\bm x,\bm x^\star)},
 \label{eq:qite_appendix_initial_probability}
\end{equation}
where the Hamming distance $d_H(\bm x,\bm x^\star)$ counts the bits that differ from the representative. The angle $\phi_r$ therefore controls how broadly the initial distribution samples the register.

For the remainder of the derivation, fix$H=H_r=\sum_U h_{U,r}Z_U$, where $h_{U,r}$ is real and  $U\subseteq\{0,\ldots,q-1\}$ specifies the support of $Z_U=\prod_{u\in U}Z_u$, with $Z_\varnothing=I$. Unless stated otherwise, brackets denote expectations in the initial state: $\langle Q\rangle=\langle\psi_{0,r}|Q|\psi_{0,r}\rangle$ for an operator $Q$. Substituting Eq.~\eqref{eq:qite_appendix_warm_state} gives the single-qubit moments
\begin{equation}
 \begin{aligned}
 \langle Z_u\rangle&=(-1)^{x_u^\star}\cos(2\phi_r),\\
 \langle X_u\rangle&=\sin(2\phi_r),
 \qquad \langle Y_u\rangle=0.
 \end{aligned}
 \label{eq:qite_single_qubit_expectations}
\end{equation}
The $X$ expectation is independent of the representative bit, while the $Z$ expectation retains its sign. Factorization across distinct qubits then gives
\begin{equation}
 \langle Z_U\rangle
 =(-1)^{\sum_{u\in U}x_u^\star}\cos^{|U|}(2\phi_r),
 \qquad \langle Z_\varnothing\rangle=1.
 \label{eq:qite_appendix_z_expectation}
\end{equation}
The empty product equals one at every angle, including $\phi_r=\pi/4$. Products of these observables also simplify because $Z_u^2=I$: repeated factors cancel, leaving $Z_UZ_\Lambda=Z_{U\triangle\Lambda}$, where the symmetric difference $U\triangle\Lambda$ contains indices present in exactly one set. Consequently, the Hamiltonian moments required for the update are
\begin{equation}
 \begin{aligned}
 \langle H\rangle&=\sum_U h_{U,r}\langle Z_U\rangle,\\
 \langle HZ_\Lambda\rangle
 &=\sum_U h_{U,r}\langle Z_{U\triangle\Lambda}\rangle.
 \end{aligned}
 \label{eq:qite_appendix_classical_moments}
\end{equation}
These moments follow entirely from the initial product state, so even contributions supported on more than two qubits can be calculated classically without circuit measurements.

\subsection{Imaginary-time target}

For the fixed Hermitian Hamiltonian $H$, normalized imaginary-time evolution is~\cite{QITE_ref}
\begin{equation}
 |\psi_{\mathrm{ITE}}(\tau)\rangle
 =\frac{e^{-\tau H}|\psi_{0,r}\rangle}
 {\|e^{-\tau H}|\psi_{0,r}\rangle\|_2},
 \qquad \tau\geq0.
 \label{eq:qite_appendix_normalized_ite}
\end{equation}
Here $\|\cdot\|_2$ is the Euclidean norm, and the imaginary-time parameter $\tau$ is dimensionless because $H$ represents a dimensionless scaled loss. Denoting expectations in the evolved state by $\langle Q\rangle_\tau$, differentiation of the exponential and its normalization gives
\begin{equation}
 \partial_\tau|\psi_{\mathrm{ITE}}(\tau)\rangle
 =-\bigl(H-\langle H\rangle_\tau I\bigr)
   |\psi_{\mathrm{ITE}}(\tau)\rangle.
 \label{eq:qite_state_derivative}
\end{equation}

We match the initial derivatives of the retained nonidentity Pauli terms in $H_r$. Let $\mathcal L_r$ contain their support sets and let $n_H=|\mathcal L_r|$ be the number of selected observables. For each $\Lambda\in\mathcal L_r$, differentiating the bra and ket of $\langle Z_\Lambda\rangle_\tau$ at $\tau=0$ yields
\begin{equation}
 \begin{aligned}
 \left.\partial_\tau\langle Z_\Lambda\rangle_\tau\right|_0
 &=-\langle HZ_\Lambda+Z_\Lambda H\rangle
   +2\langle H\rangle\langle Z_\Lambda\rangle\\
 &=2\bigl(\langle H\rangle\langle Z_\Lambda\rangle
          -\langle HZ_\Lambda\rangle\bigr)
 =2D_\Lambda.
 \end{aligned}
 \label{eq:qite_target_derivative}
\end{equation}
The second line follows from $[H,Z_\Lambda]=0$ and defines $D_\Lambda$ as half the initial expectation derivative. This target is the negative covariance of $H$ and $Z_\Lambda$ in the warm-start state, which can be evaluated by expanding the Hamiltonian:
\begin{equation}
 D_\Lambda=\sum_U h_{U,r}
 \bigl(\langle Z_U\rangle\langle Z_\Lambda\rangle
       -\langle Z_{U\triangle\Lambda}\rangle\bigr).
 \label{eq:qite_covariance_expansion}
\end{equation}
The identity contribution $U=\varnothing$ cancels in this expression, so an additive constant in $H$ has no effect on the update. Applying the same differentiation to the energy expectation gives
\begin{equation}
 \partial_\tau\langle H\rangle_\tau
 =-2\bigl(\langle H^2\rangle_\tau-\langle H\rangle_\tau^2\bigr)
 \leq0.
 \label{eq:qite_appendix_energy_descent}
\end{equation}
Exact normalized imaginary-time evolution therefore cannot increase the expected surrogate energy. The finite circuit update derived below approximates this evolution, and its error must be considered separately.

\subsection{Pair generators and response}

For $q\geq2$, use all $P=\binom q2$ qubit pairs and define
\begin{equation}
 A_{ij}=Y_iZ_j+Z_iY_j,\qquad 0\leq i<j<q.
 \label{eq:qite_appendix_generator}
\end{equation}
Each Hermitian generator acts on its two indexed qubits, with the identity on all others. Assigning one real parameter $\mu_{ij}$ to each pair gives the unitary ansatz
\begin{equation}
 \begin{aligned}
 U_{ij}(\mu_{ij})&=e^{-\mathrm i\mu_{ij}A_{ij}},\\
 U^{\mathrm{QITE}}(\bm\mu)&=\prod_{i<j}U_{ij}(\mu_{ij}),
 \qquad \bm\mu\in\mathbb R^P.
 \end{aligned}
 \label{eq:qite_appendix_circuit}
\end{equation}
The product specifies gate application in dictionary pair order, with the first applied factor appearing on the right in the matrix product. Thus, for $q=3$, the circuit applies pairs $(0,1)$, $(0,2)$, and $(1,2)$, giving $U^{\mathrm{QITE}}=U_{12}U_{02}U_{01}$.

Because the two Pauli terms within each generator commute, a pair gate factors exactly into two rotations. Using $R_Q(\theta)=e^{-\mathrm i\theta Q/2}$ for a Pauli string $Q$, we obtain
\begin{equation}
 U_{ij}(\mu_{ij})
 =R_{Y_iZ_j}(2\mu_{ij})R_{Z_iY_j}(2\mu_{ij}).
 \label{eq:qite_appendix_pair_factorization}
\end{equation}
Different pair generators need not commute, so their order affects the finite circuit even though its first derivatives at $\bm\mu=0$ are independent of that order. The QITE unitary consists of these pair rotations; the higher-locality Hamiltonian terms enter through the classical calculation of their parameters.

Define the circuit expectation $\langle Z_\Lambda\rangle_{\mathrm{QITE}}(\bm\mu)
=\langle\psi_{0,r}|U^{\mathrm{QITE}\dagger}(\bm\mu) Z_\Lambda U^{\mathrm{QITE}}(\bm\mu)|\psi_{0,r}\rangle$. Since every pair rotation is the identity at zero parameters,
\begin{equation}
 \left.\frac{\partial\langle Z_\Lambda\rangle_{\mathrm{QITE}}}
 {\partial\mu_{ij}}\right|_{\bm\mu=0}
 =\mathrm i\langle[A_{ij},Z_\Lambda]\rangle
 =2G_{\Lambda,ij},
 \label{eq:qite_unitary_derivative}
\end{equation}
Here $[A,Z]=AZ-ZA$ is the commutator. Since $[Y,Z]=2\mathrm iX$, the $Y_iZ_j$ term contributes when $i\in\Lambda$, while the $Z_iY_j$ term contributes when $j\in\Lambda$. Writing $U=\Lambda\triangle\{i,j\}$ combines these contributions as
\begin{equation}
 \begin{aligned}
 G_{\Lambda,ij}
 &=-\mathbf 1_{\{i\in\Lambda\}}\langle X_iZ_U\rangle
   -\mathbf 1_{\{j\in\Lambda\}}\langle X_jZ_U\rangle\\
 &=-|\Lambda\cap\{i,j\}|\sin(2\phi_r)\langle Z_U\rangle.
 \end{aligned}
 \label{eq:qite_response_derivation}
\end{equation}
The indicator $\mathbf 1_{\{\cdot\}}$ equals one when its condition holds and zero otherwise. For each nonzero contribution, the qubit carrying $X$ is absent from $U$, which allows the expectation to factorize. The intersection size therefore counts the contributing qubits and retains both terms when $i,j\in\Lambda$; for example, $G_{\{i,j\},ij}=-2\sin(2\phi_r)$.

Collecting the targets and responses in the same observable order gives $\bm D\in\mathbb R^{n_H}$ and $G\in\mathbb R^{n_H\times P}$. For the initial parameter velocity $\dot{\bm\mu}\in\mathbb R^P$, the target and circuit derivatives are $2\bm D$ and $2G\dot{\bm\mu}$, respectively. Matching these derivatives to first order requires
\begin{equation}
 G\dot{\bm\mu}\simeq\bm D.
 \label{eq:qite_appendix_tangent_matching}
\end{equation}
Because the chosen generators need not reproduce every target derivative, we determine the velocity through a least-squares fit.

\subsection{Regularized solve and endpoint limits}

To control large parameter velocities, we include a nonnegative regularizer $\lambda$ and solve
\begin{equation}
 \dot{\bm\mu}
 =\underset{\bm v\in\mathbb R^P}{\operatorname{arg\,min}}
 \left\{\|G\bm v-\bm D\|_2^2+\lambda\|\bm v\|_2^2\right\},
 \qquad \lambda\geq0.
 \label{eq:qite_appendix_response_solve}
\end{equation}
When $\lambda=0$ and the minimizer is not unique, we select the minimum-norm least-squares solution.

We evaluate the solution using the thin singular-value decomposition (SVD), $G=U_G\Sigma_GV_G^{\mathsf T}$. For $n_H>0$, setting $k_G=\min(n_H,P)$ gives the dimensions $U_G\in\mathbb R^{n_H\times k_G}$, $\Sigma_G\in\mathbb R^{k_G\times k_G}$, and $V_G\in\mathbb R^{P\times k_G}$. The columns of $U_G$ and $V_G$ are orthonormal, while $\Sigma_G=\operatorname{diag}(\sigma_1,\ldots,\sigma_{k_G})$ contains the nonnegative singular values in decreasing order. For $\lambda>0$, these factors give the unique solution
\begin{equation}
 \dot{\bm\mu}
 =V_G\operatorname{diag}\!\left(
 \frac{\sigma_k}{\sigma_k^2+\lambda}\right)_{k=1}^{k_G}
 U_G^{\mathsf T}\bm D.
 \label{eq:svd_filter}
\end{equation}
This expression follows by projecting the target and trial velocity onto the singular directions: $\bm\gamma=U_G^{\mathsf T}\bm D$ and $\bm\xi=V_G^{\mathsf T}\bm v$. Minimizing $(\sigma_k\xi_k-\gamma_k)^2+\lambda\xi_k^2$ along each direction gives $\xi_k=\sigma_k\gamma_k/(\sigma_k^2+\lambda)$, yielding Eq.~\eqref{eq:svd_filter} without forming $G^{\mathsf T}G$.

For $\lambda=0$, the numerical pseudoinverse replaces the filter in Eq.~\eqref{eq:svd_filter} by $1/\sigma_k$ for retained singular values and zero otherwise. The implementation retains values satisfying
\begin{equation}
 \sigma_k>\epsilon_{\mathrm{mach}}\max(n_H,P)\sigma_1,
 \label{eq:qite_appendix_svd_cutoff}
\end{equation}
where $\epsilon_{\mathrm{mach}}$ is double-precision machine epsilon. If there are no nonidentity rows or $G=0$, the minimum-norm convention selects zero velocity.

The remaining first-order mismatch is measured by the relative residual
\begin{equation}
 \varepsilon_{\mathrm{QITE}}=
 \begin{cases}
 \displaystyle\frac{\|G\dot{\bm\mu}-\bm D\|_2}{\|\bm D\|_2},
    &\|\bm D\|_2>0,\\[6pt]
 0,&\bm D=0.
 \end{cases}
 \label{eq:qite_relative_residual}
\end{equation}
Both the restricted generator family and regularization can produce a nonzero residual. A zero residual establishes agreement for the selected expectation derivatives, but this condition alone does not establish agreement for the full state derivatives.

The endpoint angles clarify the role of the warm start. At $\phi_r=0$, the initial state is the basis state $|\bm x^\star\rangle$ and hence an eigenstate of $H$. Both $\bm D$ and $G$ then vanish, so no update is selected; the algorithm therefore uses $\phi_r>0$. At $\phi_r=\pi/4$, every factor becomes $|+\rangle=(|0\rangle+|1\rangle)/\sqrt2$, independently of the representative bits. All nonempty $Z$ expectations vanish, reducing the target and response to
\begin{equation}
 D_\Lambda=-h_{\Lambda,r},\qquad
 G_{\Lambda,ij}=-2\,\mathbf 1_{\{\Lambda=\{i,j\}\}}.
 \label{eq:qite_appendix_uniform_endpoint}
\end{equation}
At this endpoint, only two-qubit $Z$ observables respond at first order, so nonzero targets with other support sizes remain unmatched. The initial register distribution is uniform, but decoding assigns probability $d(\bm z)/2^q$ to a configuration with $d(\bm z)=|\mathcal O(\bm z)|$ aliases. Thus, configurations with larger alias sets have greater initial sampling probability.

\subsection{Finite update and acquisition}

For a finite dimensionless step $\Delta\tau\geq0$, set
\begin{equation}
 \bm\mu=\Delta\tau\,\dot{\bm\mu}.
 \label{eq:qite_appendix_finite_step}
\end{equation}
For fixed $H$ and $|\psi_{0,r}\rangle$, the resulting expectation error is
\begin{equation}
 \begin{aligned}
 &\langle Z_\Lambda\rangle_{\mathrm{QITE}}(\Delta\tau\dot{\bm\mu})
 -\langle Z_\Lambda\rangle_{\tau=\Delta\tau}\\
 &\qquad=2\Delta\tau\,(G\dot{\bm\mu}-\bm D)_\Lambda
 +O(\Delta\tau^2).
 \end{aligned}
 \label{eq:qite_appendix_finite_error}
\end{equation}
The leading term is set by the matching residual, while the higher-order terms reflect the finite step beyond the initial derivative match. Increasing $\Delta\tau$ can therefore increase the approximation error, and the finite update does not guarantee a decrease in either surrogate energy or evaluated ACPF loss.

Each of the $S_r$ shots prepares Eq.~\eqref{eq:qite_appendix_warm_state}, applies Eq.~\eqref{eq:qite_appendix_circuit}, and measures the register. Classical processing converts the measured bits into action tuples, decodes their configurations, merges aliases, and excludes previously attempted configurations. The remaining candidates are ranked by the surrogate prediction averaged over each complete alias set, then submitted for ACPF evaluation. Every attempt is recorded, while only converged, finite losses enter the next fit and determine the incumbent.

To define the angle schedule, let $N$ be the usable-label cap, $n_{\mathrm{init}}\leq N$ the realized initial-label count, and $B\geq1$ the acquisition batch size. These values set the planned number of rounds to $T=\lceil(N-n_{\mathrm{init}})/B\rceil$, with the warm-start angle updated for $T>0$ as
\begin{equation}
 \phi_r=\phi_{\mathrm{start}}
 +\frac rT(\phi_{\mathrm{target}}-\phi_{\mathrm{start}}),
 \qquad r=0,\ldots,T-1,
 \label{eq:qite_appendix_outer_schedule}
\end{equation}
with $0<\phi_{\mathrm{target}}<\phi_{\mathrm{start}}\leq\pi/4$. Since $r<T$, this schedule approaches the target angle without reaching it within the stated index range; acquisition can also stop earlier when the label cap is reached or the candidate pool is exhausted. Each executed round refits $H_r$, selects a representative of the current incumbent, and prepares a fresh product state. Recalculating the analytic expectations in that state keeps the derivation valid across rounds, even though the preceding circuit can produce an entangled output.

\section{Ideal Simulation Results}
\label{app:ideal_sims}

In this appendix we present the results from 100 ideal simulations of the algorithm applied to each test case. We first address Test Cases 1--5, and then Test Cases 6--10.

\subsection{Test Cases 1-5}
We begin with the presentation of Test Cases 1-5. The experimental setup for each dataset is recapitulated in Tab.~\ref{tab:standard_case_settings_2}. The ideal simulation results are given in Tab.~\ref{tab:standard_case_results_ideal} and optimization traces are shown in Fig.~\ref{fig:case1-5_ideal_results}

\begin{widetable}
\centering
\caption{Network sizes, register allocations, and acquisition budgets
for Cases 1--5. Stage widths give the number of qubits assigned
to each branch-exchange stage.}
\label{tab:standard_case_settings_2}
\small
\setlength{\tabcolsep}{4pt}
\renewcommand{\arraystretch}{1.15}

\begin{tabular*}{\textwidth}{
    @{\extracolsep{\fill}} c c c c c c c c c c @{}
}
\toprule
& \multicolumn{3}{c}{Network}
& \multicolumn{2}{c}{Register}
& \multicolumn{4}{c}{Acquisition settings} \\
\cmidrule(lr){2-4}
\cmidrule(lr){5-6}
\cmidrule(l){7-10}

Case
& Buses
& Switches
& Ties
& Qubits
& Stage widths
& Circuits
& \shortstack{Initial ACPF\\budget}
& \shortstack{ACPF budget\\per circuit}
& \shortstack{Shots per\\circuit} \\
\midrule

1 & 14 & 16 & 3 & 11 & $(3,4,4)$
  & 19 & 1  & 1                       & 10    \\

2 & 33 & 37 & 5 & 21 & $(4,3,4,5,5)$
  & 10 & 60 & 20\textsuperscript{a}    & 2,000 \\

3 & 33 & 27 & 5 & 19 & $(2,3,4,5,5)$
  & 10 & 60 & 20\textsuperscript{a}    & 2,000 \\

4 & 69 & 73 & 5 & 26 & $(5,4,5,6,6)$
  & 10 & 60 & 44                      & 2,000 \\

5 & 69 & 54 & 5 & 21 & $(3,4,5,3,6)$
  & 10 & 60 & 44                      & 2,000 \\

\bottomrule
\end{tabular*}

\smallskip
\parbox{\textwidth}{\footnotesize
\textsuperscript{a}The ACPF budget is 20 evaluations for each of
the first nine circuits and 10 evaluations for the final circuit.}
\end{widetable}

\begin{widetable}
\centering
\caption{Radial defaults and minimal-loss configurations for Cases 1--5, specified by open-switch indices, with active losses in kW. The final column reports the median number of distinct configurations evaluated with ACPF required to reach a minimal-loss configuration.}
\label{tab:standard_case_results_ideal}
\small
\setlength{\tabcolsep}{5pt}
\renewcommand{\arraystretch}{1.15}
\begin{tabular*}{\textwidth}{
    @{\extracolsep{\fill}} c l l r r c @{}
}
\toprule
& \multicolumn{2}{c}{Open switches}
& \multicolumn{2}{c}{Loss (kW)}
& \\
\cmidrule(lr){2-3}
\cmidrule(lr){4-5}
Case
& Default
& Optimal
& Default
& Optimal
& \shortstack{ACPF evaluations\\to optimal} \\
\midrule
1 & 14, 15, 16
  & 7, 8, 16
  & 511.44 & 466.13 & 11 \\
2 & 33, 34, 35, 36, 37
  & 7, 9, 14, 32, 37
  & 202.68 & 139.55 & 172 \\
3 & \textemdash
  & \textemdash
  & \textemdash & \textemdash & 88 \\
4 & 69, 70, 71, 72, 73 
  & 14, 55, 61, 69, 70
  & 224.95 & 99.60 & 300.5 \\
5 & \textemdash
  & \textemdash
  & \textemdash & \textemdash & 220.5 \\
\bottomrule
\end{tabular*}
\end{widetable}

\subsection{Test Cases 6-10}
The experimental parameters of Cases 6--10 are given in Table \ref{tab:case_summary_2_2}. Results for the ideal simulations are presented in Tab.~\ref{tab:case_summary_4_ideal} and optimization traces are shown in Fig.~\ref{fig:rolling_traces_ideal}.

\begin{widetable}
\centering
\caption{Network sizes and rolling-window settings for Cases 6--10.
Overlap is measured in shared ties between consecutive windows.
The ACPF budgets specify the initial evaluations per window and
the additional evaluations per circuit.}
\label{tab:case_summary_2_2}
\small
\setlength{\tabcolsep}{4pt}
\renewcommand{\arraystretch}{1.15}

\begin{tabular*}{\textwidth}{
    @{\extracolsep{\fill}} c c c c c c c c c c @{}
}
\toprule
& \multicolumn{3}{c}{Network}
& \multicolumn{4}{c}{Rolling-window settings}
& \multicolumn{2}{c}{ACPF budgets} \\
\cmidrule(lr){2-4}
\cmidrule(lr){5-8}
\cmidrule(l){9-10}

Case
& Buses
& Switches
& Ties
& Windows
& \shortstack{Ties per\\window}
& \shortstack{Overlap\\(ties)}
& \shortstack{Circuits per\\window}
& \shortstack{Initial per\\window}
& \shortstack{Additional per\\circuit} \\
\midrule

6  & 70  & 79  & 11 & 5  & 4 & 2 & 6 & 40 & 20 \\
7  & 84  & 96  & 13 & 4  & 4 & 0 & 3 & 20 & 20 \\
8  & 119 & 133 & 15 & 11 & 5 & 4 & 3 & 60 & 20 \\
9  & 202 & 216 & 15 & 10 & 5 & 3 & 4 & 40 & 40 \\
10 & 415 & 473 & 59 & 15 & 4 & 0 & 2 & 40 & 30 \\

\bottomrule
\end{tabular*}
\end{widetable}

\begin{widetable}
\centering
\caption{Switch configurations and active line losses for the median of 100 ideal simulations for Cases 6--10. Configurations list the open-switch indices; ranges include both endpoints. Reference configurations and losses are taken from Ref.~\cite{9565903}.}
\label{tab:case_summary_4_ideal}
\small
\setlength{\tabcolsep}{5pt}
\renewcommand{\arraystretch}{1.12}

\begin{tabularx}{\textwidth}{
    @{} r l >{\raggedright\arraybackslash}X r @{}
}
\toprule
Case & Configuration & Open switches & Loss (kW) \\
\midrule

6 & Default
  & 69--79
  & 227.53 \\
  & Reference
  & 13, 30, 45, 51, 66, 70, 75, 76, 77, 78, 79
  & 201.41 \\
  & Final
  & 13, 28, 45, 51, 67, 70, 73, 75, 76, 78, 79
  & 203.86 \\
\addlinespace[5pt]

7 & Default
  & 84--96
  & 532.00 \\
  & Reference
  & 7, 13, 34, 39, 42, 55, 62, 72, 83, 86, 89, 90, 92
  & 469.88 \\
  & Final
  & \textemdash
  & 469.88 \\
\addlinespace[5pt]

8 & Default
  & 119--133
  & 1298.09 \\
  & Reference
  & 24, 27, 35, 40, 43, 52, 59, 72, 75, 96, 98, 110,
    123, 130, 131
  & 869.73 \\
  & Final
  & 23, 27, 35, 40, 43, 51, 62, 72, 74, 77, 83, 110, 122, 126, 131
  & 872.49 \\
\addlinespace[5pt]

9 & Default
  & 202--216
  & 548.90 \\
  & Reference
  & 12, 26, 43, 82, 118, 131, 133, 140, 168, 202, 203,
    208, 212, 213, 214
  & 511.18 \\
  & Final
  & 12, 29, 66, 74, 83, 111, 118, 125, 131, 136, 137, 140, 199, 202, 208
  & 511.61 \\
\addlinespace[5pt]

10 & Default
   & 415--473
   & 708.76 \\
   & Reference
   & 1, 2, 13, 15, 16, 30, 31, 40, 41, 50, 59, 73, 75,
     82, 94, 96, 97, 111, 115, 136, 142, 150, 155, 156,
     158, 163, 168, 169, 178, 179, 191, 195, 209, 214,
     230, 254, 256, 270, 294, 314, 317, 325, 358, 362,
     385, 389, 392, 395, 403, 404, 423, 424, 426, 436,
     437, 439, 446, 449, 466
   & 583.00 \\
   & Final
   & 1, 3, 43, 50, 51, 64, 127, 141, 165, 179, 197, 231, 277, 289, 316, 324, 327, 354, 378, 383, 407, 415, 417, 418, 419, 420, 422, 425, 426, 427, 428, 429, 430, 431, 432, 434, 435, 436, 437, 438, 440, 441, 443, 444, 445, 446, 449, 451, 453, 458, 461, 462, 464, 466, 467, 468, 470, 472, 473
   & 591.11 \\

\bottomrule
\end{tabularx}
\end{widetable}

\begin{figure*}[p]
\centering

\begin{minipage}[t]{0.48\textwidth}
    \centering
    \includegraphics[width=0.94\linewidth]{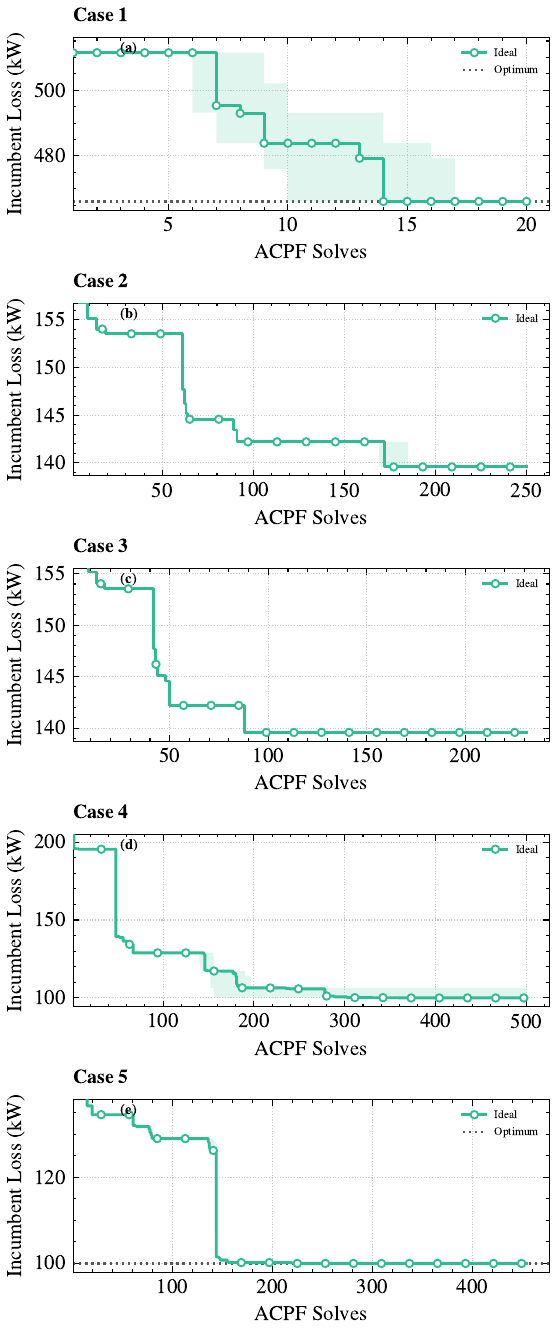}
    \caption{Best observed line loss (incumbent loss) versus cumulative ACPF solves for Cases 1--5. Curves and shaded bands give the median and interquartile range across random seeds for the ideal simulations.}
    \label{fig:case1-5_ideal_results}
\end{minipage}\hfill
\begin{minipage}[t]{0.48\textwidth}
    \centering
    \includegraphics[width=0.94\linewidth]{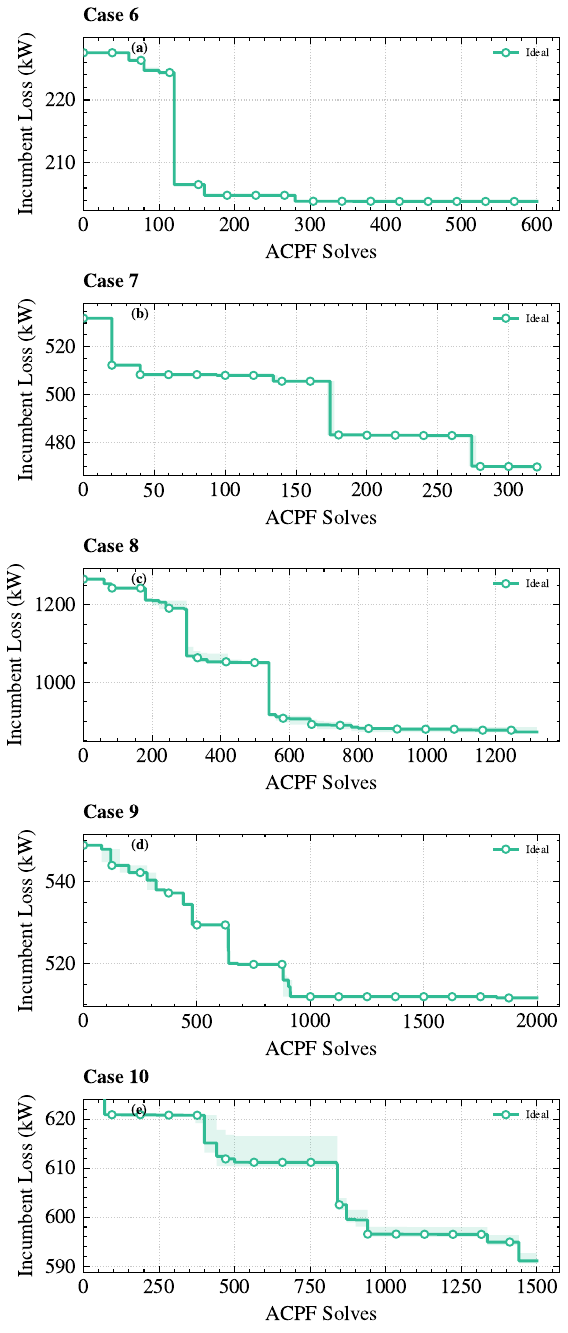}
    \caption{Best observed line loss (incumbent loss) versus cumulative ACPF solves for rolling-window Cases 6--10. Curves and shaded bands give the median and interquartile range across random seeds for the ideal simulations.}
    \label{fig:rolling_traces_ideal}
\end{minipage}

\end{figure*}

\end{document}